\documentclass{jfm}

\usepackage{graphicx}
\usepackage{natbib}

\usepackage{amsmath}
\usepackage{upmath}
\usepackage{amssymb}%
  \let\leq=\leqslant
  \let\geq=\geqslant
\usepackage{amsbsy}

\providecommand\bnabla{\boldsymbol{\nabla}}

\newcommand\Rey{\mbox{\textit{Re}}}  

\newsavebox{\astrutbox}
\sbox{\astrutbox}{\rule[-5pt]{0pt}{20pt}}

\newcommand\etal{{\it et al.}}

\newcommand\eg{e.g.}
\newcommand\ie{i.e.}

\newcommand{\bu}{{\bf u}}
\newcommand{\bU}{{\bf U}}
\newcommand{\ubar}{\langle \bu \rangle}
\newcommand{\pbar}{\langle p \rangle}

\newcommand{\bufluc}{\tilde{\bu}}
\newcommand{\ufluc}{\tilde{u}}
\newcommand{\vfluc}{\tilde{v}}

\newcommand{\la}{\langle}
\newcommand{\ra}{\rangle}
\newcommand{\bx}{{\bf x}}
\newcommand{\ex}{{\bf e_x}}
\newcommand{\ey}{{\bf e_y}}
\newcommand{\ez}{{\bf e_z}}
\newcommand{\be}{{\bf e}}

\newcommand\pd{\ensuremath{\partial}}
\newcommand{\grad}{\bnabla}
\newcommand{\lap}{\nabla^2}
\newcommand{\blap}{\bnabla^2}

\newcommand{\divv}{\bnabla \cdot}

\newcommand{\p}{^\prime}
\newcommand{\pp}{^{\prime\prime}}

\newcommand{\bg}{{\bf g}}
\newcommand{\oor}{\frac{1}{\Rey}}
\newcommand{\half}{\frac{1}{2}}
\newcommand{\ccf}{\kappa_{xyz}}
\newcommand{\cc}{\kappa_{yz}}
\newcommand{\forcevec}{\bf F}
\newcommand{\forcesca}{F}

\newcommand\comment[1]{ }

\title[Mean Flow of Turbulent-Laminar Patterns] {
Mean Flow of Turbulent-Laminar Patterns\\ in Plane Couette Flow}

\author[D. Barkley and L. S. Tuckerman]%
{D\ls W\ls I\ls G\ls H\ls T\ns B\ls A\ls R\ls K\ls L\ls E\ls Y$^1$\break
\and  
L\ls A\ls U\ls R\ls E\ls T\ls T\ls E\ns 
S.\ns T\ls U\ls C\ls K\ls E\ls R\ls M\ls A\ls N$^2$}

\affiliation{%
$^1$Mathematics Institute, University of Warwick, 
    Coventry, CV4 7AL, UK 		\\
    www.maths.warwick.ac.uk/$\sim$barkley \\
    barkley@maths.warwick.ac.uk \\
[\affilskip]
$^2$
LIMSI-CNRS, BP 133, 91403 Orsay, France \\
www.limsi.fr/Individu/laurette \\
laurette@limsi.fr
}

\pubyear{2006}
\volume{}
\pagerange{}
\date{?? and in revised form ??}

\begin{document}

\maketitle

\begin{abstract}
A turbulent-laminar banded pattern in plane Couette flow is studied
numerically. This pattern is statistically steady, is oriented obliquely to
the streamwise direction, and has a very large wavelength relative to the gap.
The mean flow, averaged in time and in the homogeneous direction, is
analysed. The flow in the quasi-laminar region is not the linear Couette
profile, but results from a non-trivial balance between advection and
diffusion.  This force balance yields a first approximation to the
relationship between the Reynolds number, angle, and wavelength of the
pattern.  Remarkably, the variation of the mean flow along the pattern
wavevector is found to be almost exactly harmonic: the flow can be represented
via only three cross-channel profiles as $\bU(x,y,z) \approx \bU_0(y) +
\bU_c(y) \cos(kz) + \bU_s(y)\sin(kz)$.  A model is formulated which relates
the cross-channel profiles of the mean flow and of the Reynolds stress.
Regimes computed for a full range of angle and Reynolds number in a tilted
rectangular periodic computational domain are presented.  Observations of
regular turbulent-laminar patterns in other shear flows -- Taylor-Couette,
rotor-stator, and plane Poiseuille -- are compared.
\end{abstract}


\section{Introduction}

Pattern formation is associated with the spontaneous breaking of spatial
symmetry.  Many of the most famous and well-studied examples of pattern
formation come from fluid dynamics. Among these are the convection rolls which
spontaneously form in a uniform layer of fluid heated from below and the
Taylor cells which form between concentric rotating cylinders.  In these cases
continuous translational symmetries are broken by the cellular flows beyond
critical values of the control parameter -- the Rayleigh number or Taylor
number.

\begin{figure}
\centerline{\hspace*{2cm}
\includegraphics[width=8cm]{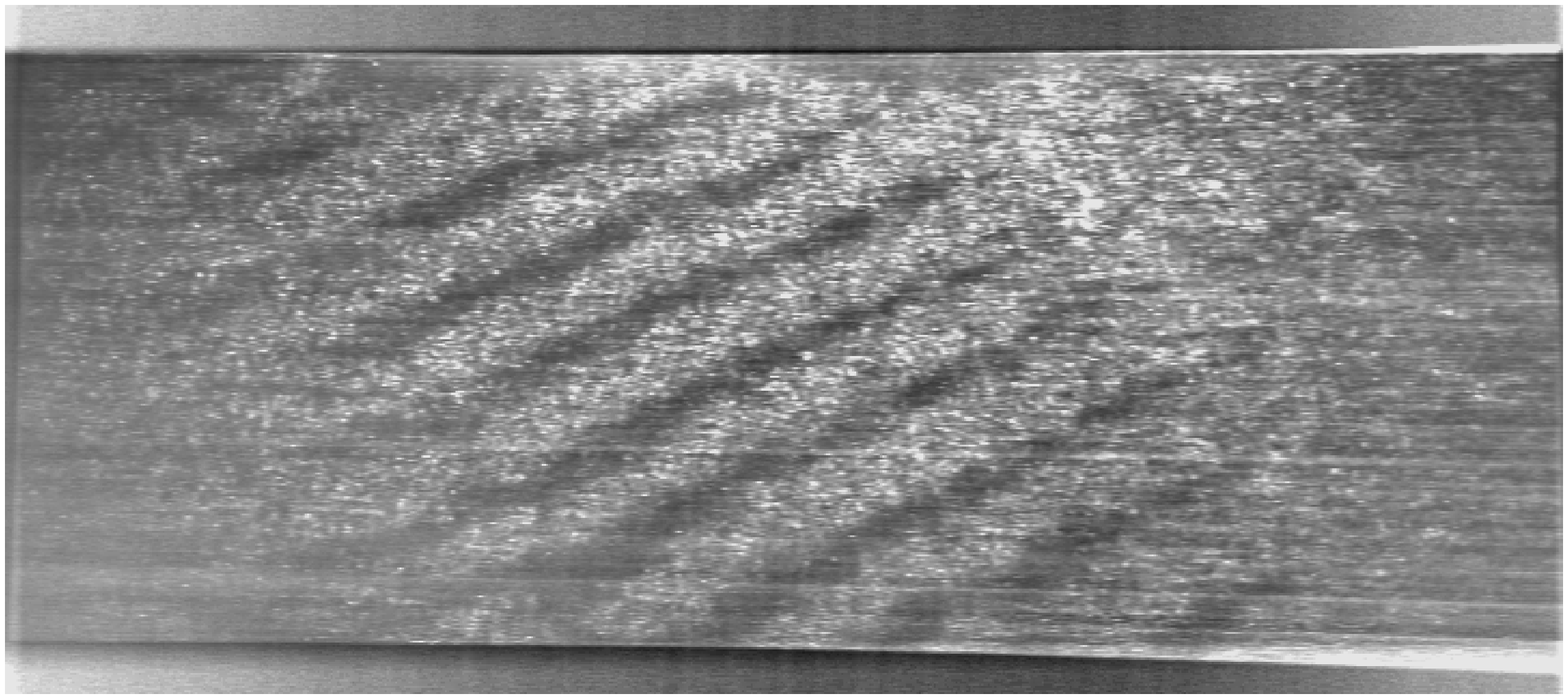}
\includegraphics[width=2.3cm]{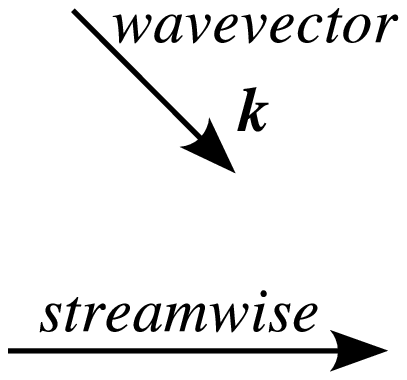}}
\caption{Photograph of a turbulent-laminar pattern in plane Couette flow from
  the Saclay experiment. Light regions correspond to turbulent flow and dark
  regions to laminar flow. The striped pattern of alternating laminar and
  turbulent flow forms with a wavevector ${\bf k}$ oblique to the streamwise
  direction. The wavelength is approximately 40 times the half-gap between the
  moving walls. The lateral dimensions are 770 by 340 half-gaps and the
  Reynolds number is $Re=385$.  Figure reproduced with permission from Prigent
  \etal}
\label{fig:experiment}
\end{figure}

A fundamentally new type of pattern has been discovered in large-aspect-ratio
shear flows in recent years by researchers at GIT-Saclay
\cite[]{Prigent_arxiv,Prigent_PRL,Prigent_PhysD,Prigent_IUTAM,Bottin_EPF}.
Figure~\ref{fig:experiment} shows an example from plane Couette experiments
performed by these researchers. One sees a remarkable spatially-periodic
pattern composed of distinct regions of turbulent and laminar flow.  The
pattern itself is essentially stationary.  The pattern wavelength is large
compared with the gap between the plates and its wavevector is oriented
obliquely to the streamwise direction.

\begin{figure}
\centerline{\includegraphics[width=8cm]{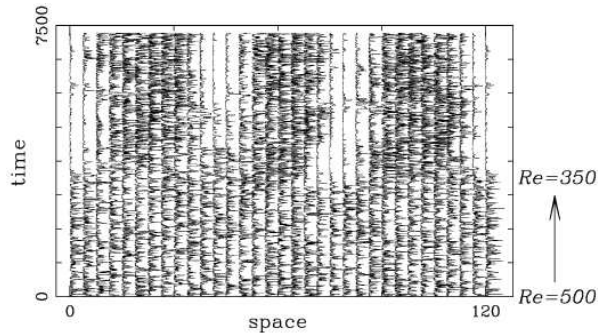}}
\caption{Space-time plot from numerical simulations of plane Couette flow
showing the spontaneous formation of a turbulent-laminar pattern at $Re=350$.
The kinetic energy in the mid-plane is sampled at 32 equally spaced points
along an oblique cut (in the direction of pattern wavevector) through three
wavelengths of the pattern.  At time zero, $Re=500$ and the flow is uniformly
turbulent. Over about 3000 time units $Re$ is decreased in steps to 350, and
then held constant.}
\label{fig:R350isbanded}
\end{figure}

The pattern emerges spontaneously from featureless turbulence as the Reynolds
number is {\em decreased}. This is illustrated in
figure~\ref{fig:R350isbanded} with time series from our numerical simulations
of plane Couette flow for decreasing Reynolds number (conventionally defined
based on half the velocity difference between the plates and half the gap).
At Reynolds number 500, the flow is uniformly turbulent. 
Following a decrease in the Reynolds number below 400 (specifically
350 in figure~\ref{fig:R350isbanded}) the flow organises into three regions of
relatively laminar flow and three regions of more strongly turbulent flow.
While the fluid in the turbulent regions is very dynamic, the pattern is
essentially steady.

Shear flows exhibiting regular coexisting turbulent and laminar regions have a
been known for many years.  In the mid 1960's, a state known as spiral
turbulence was discovered \cite[]{Coles,vanAtta,ColesvanAtta} in
counter-rotating Taylor-Couette flow. Consisting of a turbulent and a laminar
region, each with a spiral shape, spiral turbulence was further studied in the
1980s \cite*[]{Andereck,Hegseth}.  Experiments by the Saclay researchers
\cite[]{Prigent_arxiv,Prigent_PRL,Prigent_PhysD,Prigent_IUTAM} in a very large
aspect-ratio Taylor-Couette system have shown that the turbulent and laminar
regions in fact form a periodic pattern, of which the original observations of
Coles and van Atta comprised only one wavelength.  Analogues of these states
occur in other shear flows as well.  \cite{Cros} discovered large-scale
turbulent spirals in the shear flow between a stationary and a rotating disk.
\cite*{Tsukahara} observed oblique turbulent-laminar bands in plane Poiseuille
flow. A unified Reynolds number based on the shear and the half-gap can
be defined for these different flows \cite[]{Prigent_PhysD} and is described
in the Appendix.
When converted to comparable quantities in this way, the Reynolds-number
thresholds, wavelengths, and angles are similar for all of these turbulent
patterned flows.  The patterns are always found near the minimum 
Reynolds numbers for which turbulence can exist in the flow.

In this paper we present a detailed analysis of these turbulent-laminar
patterns. We will focus on a single case -- the periodic pattern at Reynolds
number 350.  From computer simulations, we obtain the flow and identify the
symmetries of the patterned state. We consider in detail the force balance
responsible for maintaining the pattern. From the symmetries and harmonic
content we are able to reduce the description to six ordinary-differential
equations which very accurately describe the patterned mean flow.

\section{Preliminaries}

\subsection{Geometry}

\begin{figure}
\centerline{
\includegraphics[width=10cm]{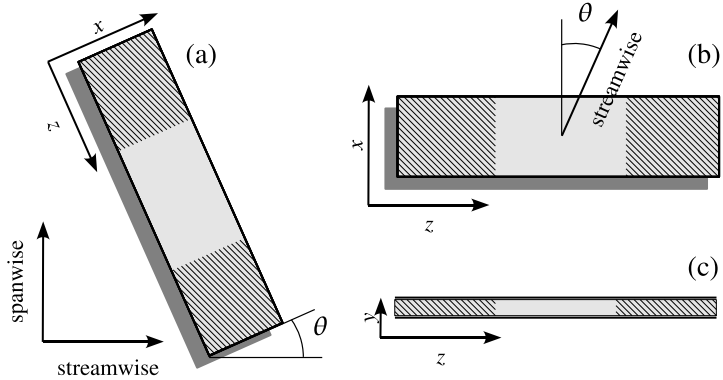}
}
\caption{Computational domain oriented at angle $\theta$ to the
streamwise-spanwise directions.  The $z$ direction is aligned to the pattern
wavevector while the $x$ direction is perpendicular to the pattern wavevector.
The turbulent region is represented schematically by hatching.
(a) Domain oriented with streamwise velocity horizontal, as in
figure~\ref{fig:experiment}.  (b) Domain oriented with $z$ horizontal, as it
will be represented in this paper.  In (a), (b) the near (upper) plate moves
in the streamwise direction; the far (lower) plate in the opposite
direction. (c) View between the plates.  }
\label{fig:domains}
\end{figure}

The unusual but key feature of our study of turbulent-laminar patterns is the
use of simulation domains aligned with the pattern wavevector and thus tilted
relative to the streamwise-spanwise directions of the flow.
Figure~\ref{fig:domains} illustrates this and defines our coordinate system.
In figure~\ref{fig:domains}(a) a simulation domain is shown as it would appear
relative to an experiment, figure \ref{fig:experiment}, in which the
streamwise direction (defined by the direction of plate motion) is horizontal.
The near (upper) plate moves to the right and the far (lower) plate to the
left in the figure.  As we have discussed in detail
\cite[]{Barkley_PRL,Barkley_IUTAM}, simulating the flow in a tilted geometry
has advantages in reducing computational expense and in facilitating the study
of pattern orientation and wavelength selection.
The important point for the present study is that the coordinates are aligned
to the patterns. The $z$ direction is parallel to the pattern wavevector while
the $x$ direction is perpendicular to the wavevector (compare
figure~\ref{fig:domains}(a) with figure~\ref{fig:experiment}).

Figures~\ref{fig:domains}(b) and (c) show the simulation domain as it will be
oriented in this paper.  In this orientation the streamwise direction is
tilted at angle $\theta$ (here $24^\circ$) to the $x$ direction.  This choice
of angle is guided by the experimental results and by our previous
simulations.  (In past publications \cite[]{Barkley_PRL,Barkley_IUTAM} we have
used un-primed $x-z$ coordinates for those aligned along spanwise-streamwise
directions and primes for coordinates tilted with the simulation domain.  Here
we focus exclusively on coordinates fixed to the simulation domain and so for
convenience denote them without primes.)  
In these tilted coordinates, the streamwise direction is
\begin{equation}
\ex\cos\theta +\ez\sin\theta \equiv \alpha \ex + \beta\ez
\label{eq:streamwise}\end{equation}
where
\begin{equation}
\alpha \equiv  \cos\theta = \cos(24^\circ)=0.913,  ~~~~~~~ 
\beta \equiv \sin\theta = \sin(24^\circ)=0.407.
\label{eq:alphabeta}
\end{equation}

We take $L_x=10$, for the reasons explained in
\cite{Jimenez,Hamilton,Waleffe_03,Barkley_PRL,Barkley_IUTAM}.  Essentially,
$L_x \sin\theta$ must be near 4 in order to contain one pair of streaks or
spanwise vortices, which are necessary to the maintenance of low Reynolds
number wall-bounded turbulence.  Although our simulations are in a
three-dimensional domain, we will average the results in the homogeneous $x$
direction, as will be explained in section \ref{sec:average}.  For most
purposes it is sufficient to view the flow in the $z-y$ coordinates
illustrated in figure~\ref{fig:domains}(c).  The midplane between the plates
corresponds to $y=0$.

The length $L_z$ of our computational domain is guided by the experimental
results and by our previous simulations.  One of the distinctive features of
the turbulent-laminar patterns is their long wavelength relative to the gap
between the plates.  A standard choice for length units in plane Couette
flow is the half-gap between the plates.  In the simulation with $L_z=120$ and
$\theta=24^\circ$ shown in figure \ref{fig:R350isbanded}, a pattern of
wavelength 40 emerged spontaneously from uniform turbulence when the Reynolds
number was lowered to $Re=350$.  For this reason, the simulations we will
describe below are conducted with $L_z=\lambda_z=40$. The corresponding
wavenumber is
\begin{equation}
k\equiv \frac{2\pi}{40}=0.157.
\end{equation}
This large wavelength, or small wavenumber, expresses the fact that the
pattern wavelength in $z$ is far greater than the cross-channel dimension.

\subsection{Equations and Numerics}
\label{sec:eqnum}

The flow is governed by the incompressible Navier--Stokes equations 
\begin{subeqnarray}
\frac{\pd \bu}{\pd t} &=& -(\bu \cdot \grad)\bu 
  - \grad p + \oor \blap \bu \quad \hbox{in $\Omega$},  \\
0&=&\divv \bu \quad\hbox{in $\Omega$},
\label{eq:nse}
\end{subeqnarray}
where $\bu(\bx,t)$ is the velocity field and $p(\bx,t)$ is the static
pressure. Without loss of generality the density is taken to be one.  The
equations have been nondimensionalized by the plate speed and the half gap
between the plates. $\Omega$ is the tilted computational domain discussed 
in the previous section.

No-slip boundary conditions are imposed at the plates and periodic boundary
conditions are imposed in the lateral directions. In our coordinates the
conditions are
\begin{subeqnarray}
\bu(x,y=\pm1,z) & = & \pm (\ex \cos\theta + \ez \sin\theta) \\ 
\bu(x + L_x,y,z) & = & \bu(x,y,z) \\
\bu(x,y,z + L_z) & = & \bu(x,y,z).
\label{eq:bcs}
\end{subeqnarray}

Linear Couette flow $\bu^L$ is a solution to \eqref{eq:nse}--\eqref{eq:bcs},
which is stable for all $\Rey$ and satisfies
\begin{equation}
\blap\bu^L = (\bu^L\cdot\grad)\bu^L = 0
\end{equation}
In our tilted coordinate system, 
\begin{equation}
\bu^L= y(\ex\cos\theta +\ez\sin\theta) = y(\alpha \ex + \beta\ez) 
= u^L\ex+w^L\ez
\label{eq:uL}
\end{equation}

The Navier-Stokes equations~\eqref{eq:nse} with boundary 
conditions~\eqref{eq:bcs}
are simulated using the spectral-element ($x$-$y$)
-- Fourier ($z$) code {\tt Prism}~\cite[]{Henderson}.  We use a spatial
resolution consistent with previous studies
\cite[]{Hamilton,Waleffe_03}. Specifically, for a domain with dimensions $L_x=10$ 
and $L_y=2$, we use a computational grid with 10 elements in the $x$
direction and 5 elements in the $y$ direction.  Within each element, we
use $8$th order polynomial expansions for the primitive variables.  In
the $z$ direction, a Fourier representation is used and the code is
parallelized over the Fourier modes.  Our domain with
$L_z=40$ is discretized with 512 Fourier modes or
gridpoints. Thus the total spatial resolution we use for the $L_x\times L_y
\times L_z = 10\times 2\times 40$ domain can be expressed as $N_x \times
N_{y} \times N_z = 81 \times 41 \times 512 = 1.7\times 10^6$ modes or 
gridpoints.

\subsection{Dataset and averaging}
\label{sec:average}

\begin{figure}
\centerline{
\includegraphics[width=14cm]{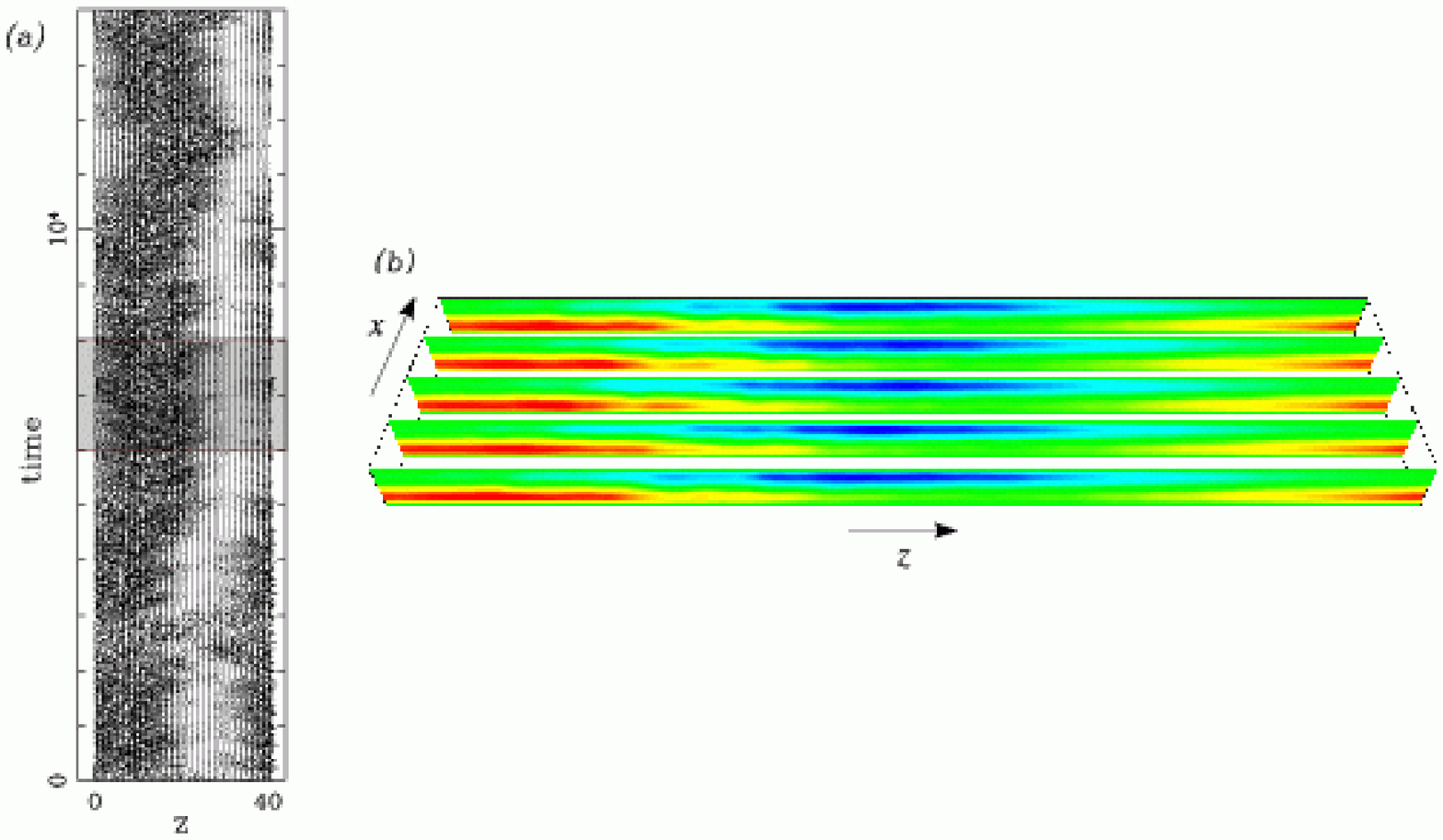} }
\caption{(a) Timeseries of a turbulent-laminar pattern.  Shown is the kinetic
  energy $E=\bu\cdot\bu/2$ along the line $x=y=0$ at 32 equally spaced points
  in $z$ for $0\leq t \leq 14 000$.  The interval [6000,8000] used for time
  averaging is shown in grey.
  (b) Time-averaged velocity at five $x$ locations illustrating the
  $x$-independence of the time-averaged flow.  Plotted is $\frac{1}{T}\int dt
  (u - u^L)$, the average $x$ component of velocity with linear Couette flow
  subtracted, averaged over the interval [6000,8000] indicated in (a).  Color
  range from blue to red: [--0.4, 0.4].}
\label{fig:average}

\centerline{
\includegraphics[width=14cm]{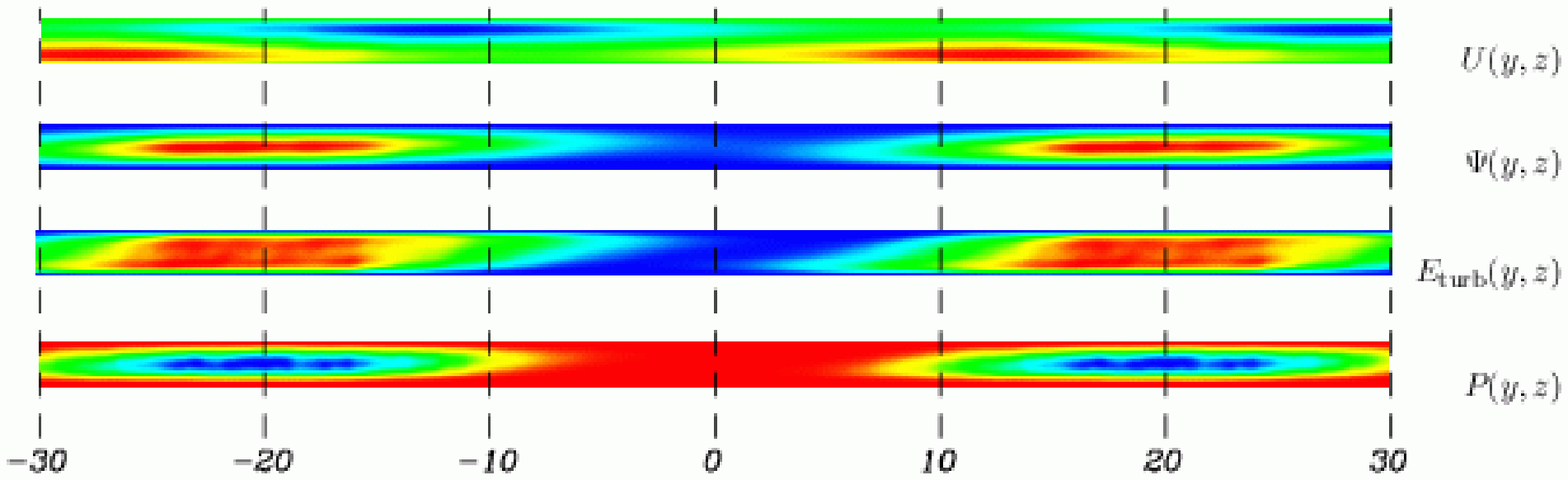}}
\caption{ $U(y,z)$: transverse component of mean flow.  $\Psi(y,z)$:
streamfunction of in-plane mean flow.  A long cell extends from one
laminar-turbulent boundary to the other.  Gradients of $\Psi$ are much larger
in $y$ than in $z$, \ie\ $|W|\gg|V|$.  In the laminar region at the center,
$W, V\approx 0$.  $E_{\rm turb}(y,z)$: mean turbulent kinetic energy $\langle
\bufluc\cdot\bufluc\rangle/2$.  There is a phase difference of
$\lambda_z/4=10$ between extrema of $E_{\rm turb}$ and of $U$.  $P(y,z)$: mean
pressure field.  Pressure gradients are primarily in the $y$ direction and
within the turbulent region.  Color ranges for each field from blue to red:
$U$ [--0.4, 0.4], $\Psi$ [0, 0.09], $E_{\rm turb}$ [0, 0.4], P [0, 0.007] .}
\label{fig:upsikep}
\end{figure}

The focus of this paper is the mean field calculated from the simulation
illustrated by the spatio-temporal diagram in figure \ref{fig:average}(a).
The velocity field in the portion of the domain shows high-frequency and
high-amplitude fluctuations, while the flow in the right portion is basically
quiescent.  We will call the flow on the left turbulent, even though it could
be argued that it is not fully developped turbulence.  We will call the flow
on the right laminar, even though occasional small fluctuations can be seen in
this region.

The turbulent-laminar pattern subsists during the entire simulation of $14
\times 10^3$ time units.  However the pattern undergoes short-scale
``jiggling'', seen particularly at the edges of the turbulent regions, and
longer-scale drifting or wandering in the periodic $z$ direction.  We seek to
describe the field which results from smoothing the turbulent fluctuations,
but for which drifting is minimal, by averaging over an appropriate time
interval.  The desired averaging time interval represents a compromise between
the short and long timescales.  We have chosen to average the flow in figure
\ref{fig:average}(a) over the shaded time interval $[t,t+T]=[6000,8000]$,
during which the pattern is approximately stationary.

The time-averaged flow is homogeneous in $x$-direction. This is illustrated in
figure~\ref{fig:average}(b) where we plot one of the velocity components
time-averaged flow over the interval $[6000,8000]$.  Cuts at different $x$
locations show that there is essentially no variation in the $x$-direction.
All other quantities are similarly independent of $x$.  It is therefore
appropriate to consider mean flows as averages over the $x$ direction as well
as over the time.

We define mean flows as
\begin{subeqnarray}
\ubar(y,z) & \equiv & \frac{1}{T} \frac{1}{L_x} \int_t^{t+T} 
\int_0^{L_x} \bu(x,y,z,t) \; dx dt \\ 
\pbar(y,z) &\equiv& \frac{1}{T} \frac{1}{L_x} \int_t^{t+T} 
\int_0^{L_x} p(x,y,z,t) \; dx dt. 
\end{subeqnarray}
The mean fields obey the averaged Navier-Stokes equations
\begin{subeqnarray}
\label{eq:NSavg}
0&=&-\left(\ubar\cdot \grad \right) \ubar 
- \la \left( \bufluc \cdot \grad \right) \bufluc \ra
-\grad \pbar + \oor \blap \ubar \\
0&=&\divv \ubar,
\end{subeqnarray}
where
\begin{equation}
\bufluc \equiv \bu - \ubar 
\end{equation}
is the fluctuating field and $\la \ra$ denotes $x$-$t$ average.  The mean
fields are subject to the same boundary conditions as equations~\eqref{eq:nse}.
We denote the Reynolds-stress force from the fluctuating field in
equations~\eqref{eq:NSavg} by $\forcevec$:
\begin{equation}
\forcevec \equiv - \la\left(\bufluc \cdot \grad \right) \bufluc \ra 
= - \divv \la \bufluc \bufluc \ra,
\end{equation}
We shall focus almost exclusively on the difference between the mean flow and
linear Couette flow, for which we introduce the notation
\begin{equation}
\bU \equiv \ubar - \bu^L,
\label{eq:UminusC}
\end{equation}
as well as $P\equiv \la p \ra$.

Letting the components of $\bU$ be denoted by $(U,V,W)$ and the components of
$\forcevec$ be denoted by $(\forcesca^U,\forcesca^V,\forcesca^W)$, then the averaged Navier-Stokes equations
for the deviation from linear Couette flow in component form become
\begin{subequations}
\label{eq:RANS}
\begin{eqnarray}
0 &=& -\left(V \pd_y + (W + \beta y)~ \pd_z \right) (U + \alpha y) 
~~~~~~~~~~+ \oor (\pd^2_y + \pd^2_z) U + \forcesca^U \label{eq:NSavgU}  \\
0 &=& -\left(V \pd_y + (W + \beta y)~ \pd_z  \right) V ~~~~~~~~~~
 -\pd_y P + \oor (\pd^2_y + \pd^2_z) V + \forcesca^V \label{eq:NSavgV}\\
0 &=& -\left(V \pd_y + (W + \beta y)~ \pd_z \right) (W + \beta y) 
-\pd_z P + \oor (\pd^2_y + \pd^2_z) W + \forcesca^W \label{eq:NSavgW}
\end{eqnarray}
\begin{eqnarray}
0 &=& \pd_y V + \pd_z W. \label{eq:RANSdiv}
\end{eqnarray}
\end{subequations}
$\bU$ is required to satisfy homogeneous boundary conditions at the plates
\begin{equation}
\bU(y=\pm1,z) = {\bf 0}
\label{eq:homogbc}\end{equation} 
and periodic boundary conditions in $z$. 

A system of this type, with three components depending on two coordinates, 
is sometimes called 2.5 dimensional.
The transverse, or out-of-plane flow $U(y,z)$ appears only in the first
equation and is effectively a passive scalar advected by the in-plane flow
$(V,W)$ and driven by the Reynolds-stress force $\forcesca^U$.
The in-plane flow can be expressed in terms of a streamfunction $\Psi$ where
\begin{equation}
V\ey+W\ez=\ex \times \nabla \Psi = 
-\pd_z \Psi \ey + \pd_y \Psi \ez.
\label{eq:definepsi}\end{equation}
We shall use both $(U,V,W)(y,z)$ and $(U,\Psi)(y,z)$ to describe
the mean flows.

\comment{Should we re-write the equations using the streamfunction?
Only if boundary layer theory makes it easy. What about Reynolds-stress force?
Should it be F for consistency? Words: Reynolds-stress vs. turbulent force.}

\comment{Something somewhere about energy of mean flow and energy of
fluctuations.  Show drift vs smoothness and
averaging and stationarity.}

\section{Results}

We present a characterisation of the turbulent-laminar pattern at
$\Rey=350$. We describe in detail the mean flow, its symmetries, and the
dominant force balances within the flow.
Our goal here is not to consider closures for averaged Navier-Stokes equations
(\ref{eq:RANS}).  We will make no attempt to model the turbulence, \ie\ to
relate the Reynolds-stress tensor $\la \bufluc \bufluc \ra$ to the mean flow
$\bU$.  Instead we use fully resolved (three-dimensional, time-dependent)
numerical simulations of the turbulent flow to measure both the mean field
$\bU$ and Reynolds-stress force $\forcevec$.
From these we extract the structure of these fields and the dominant force
balances at play in sustaining turbulent-laminar patterns.

\subsection{Mean flow}
\label{sec:meanflow}

The mean flow is visualised in figure~\ref{fig:upsikep} via the transverse,
out-of-plane flow $U(y,z)$ and the in-plane streamfunction $\Psi(y,z)$.
Recall [equation \eqref{eq:UminusC}] that these fields are the deviations of
the mean flow from linear Couette flow $\bu^L$.  The mean turbulent kinetic
energy
\begin{equation}
E_{\rm turb}\equiv\frac{1}{2}\langle \bufluc\cdot\bufluc\rangle
\end{equation}
serves to clearly identify the turbulent region.  In these and subsequent
plots, the middle of the laminar region is positioned at the centre of the
figure and the turbulent region at the periodic boundaries of the
computational domain.  In figure~\ref{fig:upsikep} (but not in subsequent
figures), plots are extended in the $z$-direction one quarter-period beyond
each periodic boundary to help visualise the flow in the turbulent region.
The pattern wavelength is $\lambda_z=40$, so that $z=30$ and $z=-10$ describe
the same point, as do $z=-30$ and $z=10$.

The mean flow can be described as follows.  $U$ is strongest in the
turbulent-laminar transition regions.  In the transition region to the left of
centre ($z=-10$) in figure \ref{fig:upsikep}, $U$ is negative and primarily
in the upper half of the channel.  To the right of centre ($z=10$), $U$ is
positive and is seen primarily in the lower half of the channel.
Comparison with turbulent kinetic energy shows that the transverse mean flow
$U$ is out of phase with respect to the fluctuating field $\bufluc$ by
$\lambda_z/4$.  This has been seen experimentally by \cite{ColesvanAtta} and
Prigent \etal\ (\citeyear{Prigent_PRL}, \citeyear{Prigent_PhysD},
\citeyear{Prigent_IUTAM}).

The in-plane flow $\Psi$ in figure \ref{fig:upsikep} has a large-aspect
ratio cellular structure consisting of alternating elliptical and hyperbolic
points. The flow around the elliptical points, located in the centre of the
turbulent regions, rotates in a counter-clockwise sense, opposing linear
Couette flow.  In the vicinity of the hyperbolic points, centred in the
laminar regions, the in-plane deviation from linear Couette flow is very weak
($W$ and $V$ nearly zero).

\begin{figure}
\centerline{
\includegraphics[width=14cm]{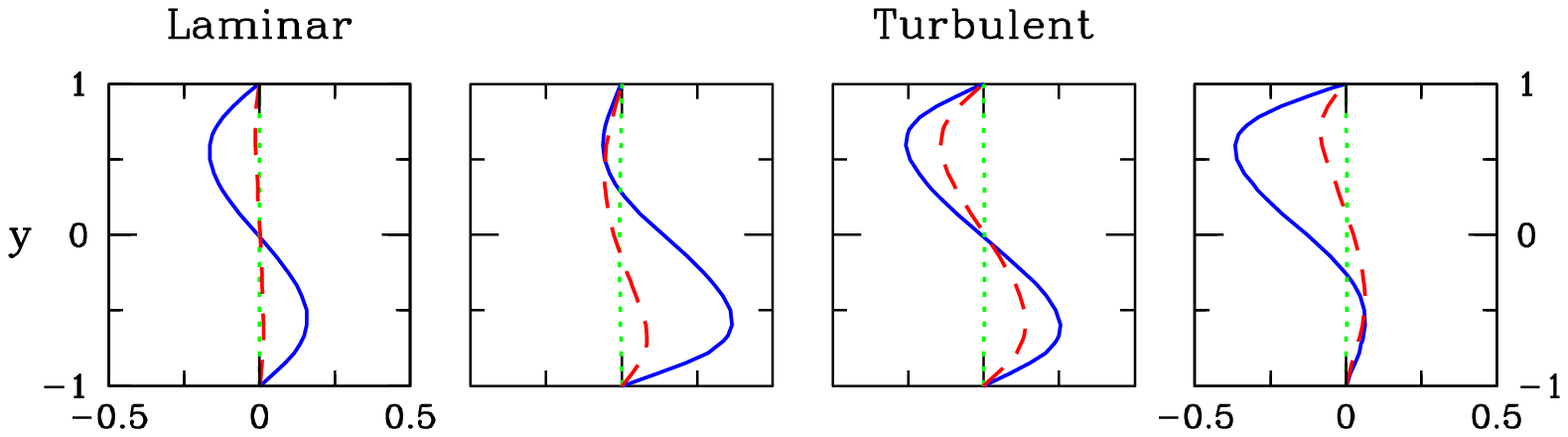}}
\caption{Mean flow profiles in $y$ at four equally spaced locations in $z$.
  From left to right: centre of the laminar region ($z=0$), laminar-turbulent
  boundary ($z=10$), centre of the turbulent region ($z=20$) and
  turbulent-laminar boundary ($z=-10$).  Components $U$ (blue, solid), $V$
  (green, dotted), $W$ (red, dashed) of deviation from linear Couette flow
  $\bu^L$.  In the laminar region, $W\approx 0$, indicating no deviation from
  $u^L$. $V$ is very small throughout.}
\label{fig:meanflow}
%
\centerline{
\includegraphics[width=14cm]{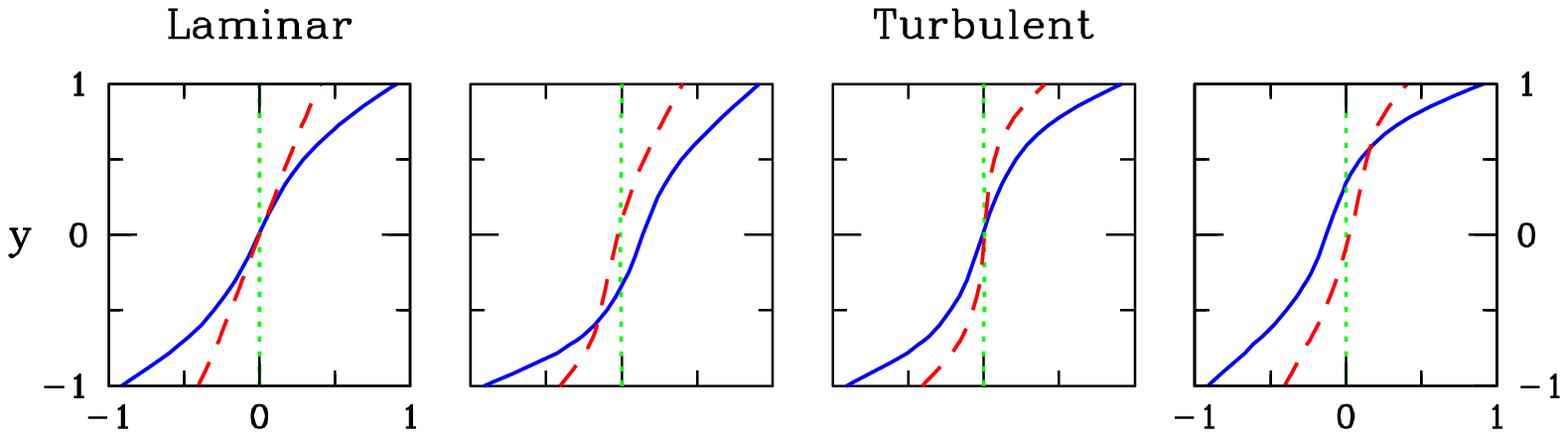}}
\caption{ Same as in figure~\ref{fig:meanflow}, but with laminar Couette flow
$\bu^L$ included.}
\label{fig:meanflowwithCou}
\end{figure}

Figure~\ref{fig:meanflow} shows $y$-profiles at four key points equally spaced
along the pattern: centre of the laminar region, turbulent-laminar transition
region, centre of the turbulent region, and the other turbulent-laminar
transition region.  While the $V$ profile is plotted, its variation is very
small on the scale of $U$ and $W$ and can essentially be used to indicate the
axis.  Figure~\ref{fig:meanflowwithCou} shows profiles for the full mean flow
$\ubar = \bU + \bu^L$ containing the linear Couette profile.

\begin{figure}
\centerline{
\includegraphics[width=10cm]{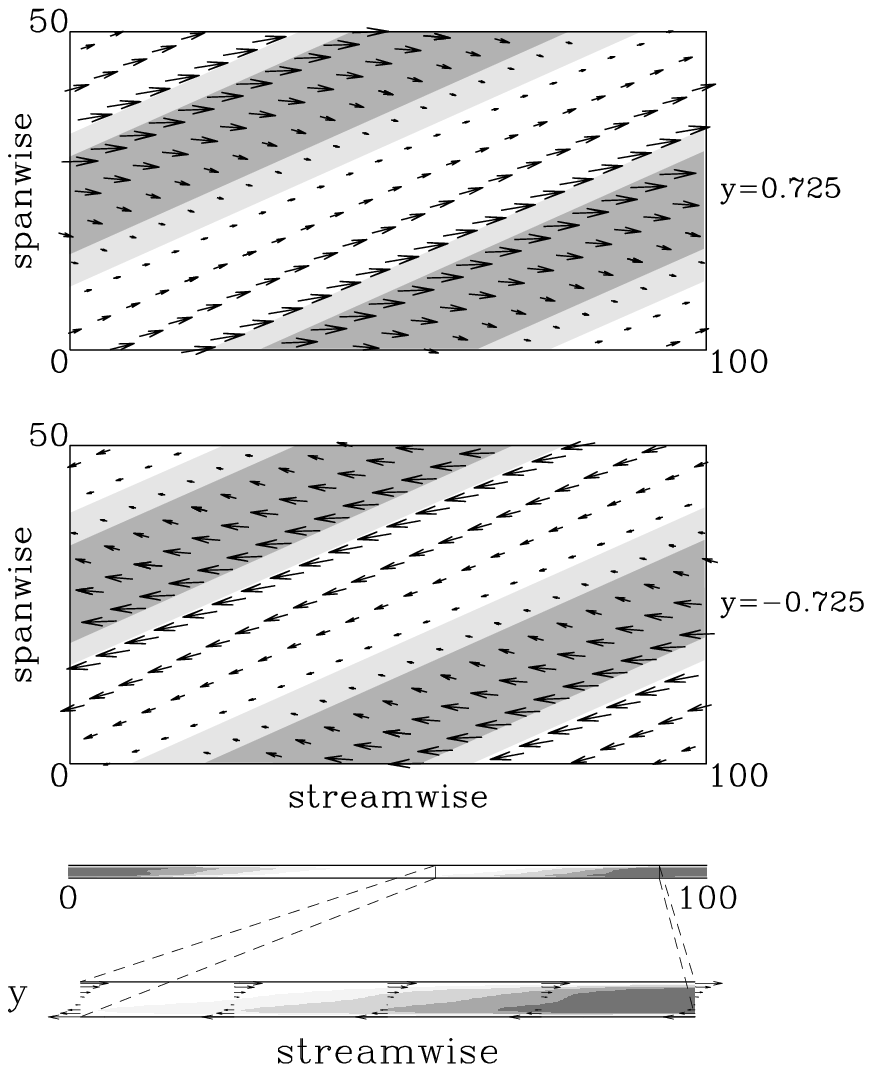}  }
\caption{ Mean velocity components seen in three planes with standard
orientation for Couette flow.  The turbulent regions are shaded.  Top:
velocity components in the streamwise-spanwise plane at $y=0.725$ (upper part
of the channel). Middle: same except $y=-0.725$ (lower part of the channel).
Bottom: flow in a constant spanwise cut. The mean velocity is shown in the
enlarged region.  }
\label{fig:coles}
\end{figure}

The $U$ profiles in figure \ref{fig:meanflowwithCou} are S-shaped, of the type
found in turbulent Couette flow. This is to be expected in the turbulent
region, even at these low Reynolds numbers.  However, it is very surprising
that the $U$ profile in the laminar region is also of this form.  In the
laminar region, local Reynolds stresses are absent (see
figure~\ref{fig:upsikep}) and so cannot be responsible for maintaining the
S-shaped velocity profile in the laminar regions.  The other prominent
features in figures~\ref{fig:meanflow} and \ref{fig:meanflowwithCou} are the
asymmetric profiles at the transition regions.

The relationship between the mean flow field and the regions of turbulence can
be seen in figure~\ref{fig:coles}. Here the flow is shown in the standard
orientations. In each view,
greyscale indicates the size of the turbulent energy and the arrows show the
mean flow within the plane.  In the top two views, the flow is shown in the
streamwise-spanwise planes located at $y=0.725$ and at $y=-0.725$.  The next
view shows the flow between the plates, \ie in a 
streamwise-cross-channel plane, and the last shows an enlargement of
one of the laminar-turbulent transition regions.
Note that the length $L_z=40$ of our tilted computational domain
corresponds to a streamwise length of $L_z/\sin\theta = 40/.407=98.3\approx
100$ and to a spanwise length of $L_z/\cos\theta= 43.78$.  

The flow in figure~\ref{fig:coles} can be compared with the mean flow reported
by \cite{ColesvanAtta} in experiments on turbulent spirals in Taylor-Coutte
flow. Coles and van Atta measured the mean flow near the midgap between the
rotating cylinders and noted an asymmetry between the mean flow into and out
of turbulent regions. They found that the mean flow into turbulent regions was
almost perpendicular to the turbulent-laminar interface whereas flow out of
the turbulent region was almost parallel to the turbulent-laminar interface.
We also observe a striking asymmetry between the mean flow into and out of the
turbulent regions. The orientation of our mean flow does not agree in detail
with that of Coles and van Atta, but this is most likely due to the fact that
Coles and van Atta considered circular Taylor-Couette flow and measured the
flow near the mid-gap.
Referring to figures~\ref{fig:upsikep} and \ref{fig:meanflow} one sees that
the mid-plane ($y=0$) is not the ideal plane on which to obverve the mean flow
since its structure is most pronounced between the midplane and the upper or
lower walls.

\begin{figure}
\centerline{
\hspace*{-1cm}
\includegraphics[width=13cm]{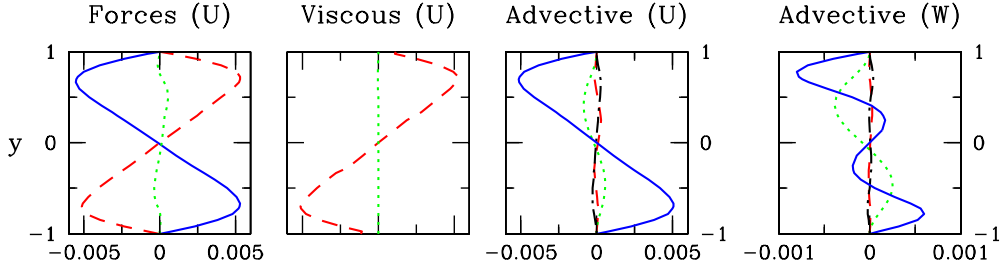}}
\caption{Balance of forces in the center of the laminar region.  Left: Forces
in the $U$ direction. Advective terms (blue, solid), viscous terms (red,
dashed), Reynolds-stress terms (green, dotted).  Middle: Viscous terms in the
$U$ direction. $(1/Re)\,\pd^2_y U$ (red, dashed) dominates dominates
$(1/Re)\,\pd^2_z U$ (green, dotted).  Right: Advective terms in the $U$
direction.  Curves show $-\beta\,y\,\pd_z U$ (blue, solid) $-W\pd_z U$ (black,
dash-dot), $-\alpha V$ (green, dotted), $-V\pd_y U$ (red, dashed).
Right-most: Advective terms in the $W$ direction (for later reference).
Curves show $-\beta\, y\, \pd_z W$ (blue, solid) $-W\pd_z W$ (black,
dash-dot), $-\beta V$ (green, dotted), $-V\pd_y W$ (red, dashed).
}
\label{fig:forcepreface}
\end{figure}

Before considering the symmetries and force balances in detail, it is
instructive to consider the dominant force balance just at the centre of the
laminar region.  
Recall that one of the more interesting features of the mean flow is that the
$U$ profile appears very similar to a turbulent profile, even in the absence
of turbulence in the laminar region.
Here the balance is dominated by advection and viscous diffusion, as
shown in figure \ref{fig:forcepreface}. Equation \eqref{eq:NSavgU} for flow in
the $x$-direction is
\begin{equation}
0 = -\left(V \pd_y + (W  + \beta y) \pd_z \right) (U + \alpha y)
 + \oor (\pd^2_y + \pd^2_z) U + \forcesca^U. 
\end{equation}

Variations in $y$ dominate variations in $z$, \ie\ the usual boundary-layer
approximation $(\pd^2_y + \pd^2_z) U \simeq \pd^2_y U$ holds; 
see, \eg\, \cite{Pope}. 
Indeed, approximating the
$y$ dependence of $U$ by the functional form $\sin(\pi y)$
suggested by figure \ref{fig:meanflow}, we have
\begin{equation}
O\left(\frac{\pd^2_y U}{\pd^2_z U}\right) = 
\frac{\pi^2}{k^2} = \frac{\pi^2}{(2\pi/40)^2}=400
\label{eq:d2ytod2z}\end{equation}
This is confirmed by the second panel of figure \ref{fig:forcepreface}.
In the centre of the laminar region $\forcesca^U$, $V$, and $W$ are all negligible,
so that $-\beta y \pd_z U$ dominates the advective terms, as shown in
the third panel of figure \ref{fig:forcepreface}.
Thus the balance between advection and viscosity in the laminar region is
\begin{equation}
\beta\: y \: \pd_z U \approx \oor \pd^2_y U.
\label{eq:firstapprox}
\end{equation}

This equation is appealingly simple and yet leads immediately to some
interesting conclusions. The first is that a non-zero tilt angle $\theta$ is
necessary to maintain the S-shaped $U$ profile in the laminar region, since
otherwise $\beta = \sin \theta = 0$ and $U$ could be at most linear in $y$ and
would in fact be zero, due to the homogeneous boundary conditions
(\ref{eq:homogbc}).  The second conclusion follows from consideration of $y$
parity.  The multiplication by $y$ on the left-hand-side reverses $y$-parity,
while the second derivative operator on the right-hand-side preserves $y$
parity.  The conclusion is that $U$ should be decomposed into odd and even
components in $y$ and equation~\eqref{eq:firstapprox} should actually be
understood as two equations coupling the two components.  Specifically, as can
be seen in figure~\ref{fig:meanflow}, $U$ is odd in $y$ in the centre of the
laminar region, yet $\pd_z U$ must be even for equation \eqref{eq:firstapprox}
to hold.  
  
The remainder of the paper is devoted to formalising, demonstrating 
and extending this basic idea.

\comment{$-\beta V$ is about half the size of $-\beta y \pd_z W$.
Maybe it should be included in the simplified model after all.
Or maybe they are all small.}
\subsection{Symmetry and Fourier modes}
\label{sec:Symmetry}

We now consider in depth the symmetry properties of the flow.  
We start with the symmetries of the system before averaging, that is, the
Navier-Stokes equations \eqref{eq:nse} and boundary conditions \eqref{eq:bcs}.
The system has translation symmetry in $x$ and $z$ as well as centrosymmetry
under combined reflection in $x$, $y$ and $z$:
\begin{equation}
\ccf(u,v,w)(x,y,z) \equiv (-u,-v,-w)(x_0-x,-y,z_0-z)
\label{eq:Cfdef}
\end{equation}
where the origin $x_0$, $z_0$ can be chosen arbitrarily.  Linear Couette
flow $\bu^L$ possesses all the system symmetries, as does the mean flow at
Reynolds numbers for which turbulence is statistically homogeneous in $x$ and
$z$.

Note that in the absence of tilt ($\theta=0$), the system possesses two
reflection symmetries. These can be taken to be $\ccf$ and reflection in the
spanwise direction.  For the tilted domain (at angles other than multiples of
$90^\circ$), the only reflection symmetry is $\ccf$. This can be seen in
figure~\ref{fig:domains}(a): for general tilt angles $\theta$, spanwise
reflection does not preserve the domain, \ie\ does not leave the periodic
boundaries in place.
The experimental system shown in figure~\ref{fig:experiment} possesses
spanwise reflection symmetry and hence bands can be observed in the either of
the two symmetrically related angles, the choice is dictated by factor such as
initial conditions.
By design, our tiled computational domain precludes the symmetry-related
pattern given by spanwise reflection.

The transition to the turbulent-laminar patterned state breaks symmetry.
Specifically, both the mean flow and the Reynolds-stress force break
$z$-translation symmetry but break neither $x$-translation symmetry nor
centrosymmetry. The spatial phase of the pattern in $z$ is arbitrary, but
given a phase there are two values of $z_0$, separated by half a period, for
which the flow is invariant under $\ccf$, as is typical for a circle pitchfork
bifurcation \cite*[]{Crawford}.  As can be seen in figure~\ref{fig:upsikep},
the values of $z_0$ about which the patterns are centrosymmetric are the
centres of the laminar ($z_0=0$) and of the turbulent ($z_0=\pm 20$) regions.

The centrosymmetry operator for our averaged fields $\bU$, which depend only
on $y$ and $z$, is
\begin{equation}
\cc(U,V,W)(y,z) \equiv (-U,-V,-W)(-y,z_0-z)
\label{eq:Cdef}
\end{equation}
Since the Reynolds-stress force $(\forcesca^U, \forcesca^V, \forcesca^W)$ is centrosymmetric in the
case we consider, then the averaged equations~\eqref{eq:RANS} for the mean
field have centrosymmetry. 

We formalise this further as follows.  Any $x$-independent field $\bg$ can be
decomposed into even and odd functions of $y$ and $z$ as
\begin{equation}
\bg(y,z)=\bg_{oo}(y,z) + \bg_{oe}(y,z) + \bg_{eo}(y,z) + \bg_{ee}(y,z)
\label{eq:4fcns}
\end{equation}
where, for example, $\bg_{oe}$ is odd in $y$ and even in $z-z_0$.  Applying the
operator in (\ref{eq:Cdef}) to (\ref{eq:4fcns}), we obtain
\begin{eqnarray}
\cc\bg(y,z)
&=&-\bg_{oo}(-y,z_0-z) - \bg_{oe}(-y,z_0-z) 
- \bg_{eo}(-y,z_0-z) - \bg_{ee}(-y,z_0-z)
\nonumber\\
&=&-\bg_{oo}(y,z) + \bg_{oe}(y,z) + \bg_{eo}(y,z) - \bg_{ee}(y,z)
\end{eqnarray}
For the field $\bg$ to be centrosymmetric requires $\cc\bg = \bg$, so that in
fact
\begin{equation}
\bg(y,z)=\bg_{oe}(y,z) + \bg_{eo}(y,z)
\label{eq:2fcns}
\end{equation}
Table \ref{tab:symmetry}, as well as figure~\ref{fig:upsikep}, shows that this
is indeed the case for $\bU$; it holds for $\forcevec$ as well.

\comment{Maybe SM figure showing symmetric/antisymmetric energy? Reference to
other turbulent bifs and statistical symmetry.}

\renewcommand{\arraystretch}{2}
\begin{table}
\begin{eqnarray}
\begin{array}{c||c|c|} 
&  {\mbox{$z$ even}} & {\mbox{$z$ odd}} \\ 
\cline{1-3}
{\mbox{$y$ even}} & {0.03\%} & {25.48\%} \\ 
\cline{1-3}
{\mbox{$y$ odd}} & {74.48\%} & {0.01\%} \\ 
\cline{1-3}
\end{array}
\nonumber\end{eqnarray}
\caption{
Energy $\int dx\int dy\int dz \:|\bU|^2/2$ of deviation from Couette flow 
contained in modes with different symmetries in $y$ and $z$. 
  Modes with centrosymmetry (opposite parity in $y$ and $z$) contain
  74.48\%+25.48\%=99.96\% of the total energy.  Reflection in $z$ is about the
  centre of the laminar region.}
\label{tab:symmetry}\end{table}

\begin{table}
\begin{eqnarray}
\begin{array}{c||ccc|cc|} 
& \multicolumn{3}{|c|} {\mbox{$z$ even (cosine)}} & 
\multicolumn{2}{|c|} {\mbox{$z$ odd (sine)}}\\ 
\mbox{$z$ wavenumber}& 0 & k & \geq 2k & k & \geq 2k \\ \cline{1-6}
\mbox{$y$ even} &&&&\fbox{25.2\%} & {0.3\%} \\ \cline{1-6}
\mbox{$y$ odd} & \fbox{69.7\%} & \fbox{4.7\%} & 0.1\% & & \\ \cline{1-6}
\end{array}
\nonumber\end{eqnarray}
\caption{
Energy contained in $z$ Fourier modes.
Modes retained (in boxes) are $\bU_0(y)$, $\bU_c(y)\cos(kz)$ and
$\bU_s(y)\sin(kz)$. These contain 69.7\%+4.7\%+25.2\%=99.6\%
of the total energy.}
\label{tab:fourier}
\end{table}

We now Fourier transform in $z$ to further decompose the mean velocity and
the Reynolds-stress force.  We find that the $z$-wavenumbers
$0$ and $\pm k$ have contributions to $\bU$ which
are an order of magnitude higher than the remaining wavenumber combinations.
See table~\ref{tab:fourier}. 
The deviation from the $z$ average is
thus almost exactly trigonometric, with almost no higher harmonic content.  
The dominance of these terms in the Fourier series means that $\bU$ and
$\forcevec$ can be represented by only three functions of $y$, namely:
\begin{equation}
\bg(x,y,z)=\bg_0(y) + \bg_c(y) \cos(kz) + \bg_s(y) \sin(kz)
\label{eq:3fcns}
\end{equation}
which is a special case of \eqref{eq:2fcns}, with the first two terms of
\eqref{eq:3fcns} coinciding with $\bg_{oe}(y,z)$ and the last to $\bg_{eo}$.
Thus, $\bg_0$ and $\bg_c$ are odd functions of $y$, while $\bg_s$ is even.
The fields thus consist of a $z$-independent component $\bg_0$
and two components which vary trigonometrically and out of
phase with one another, $\bg_c$ dominating in the laminar and turbulent
regions and $\bg_s$ dominating in the boundaries between them. Moreover,
$\bg_s$ dominates in the bulk, since $\bg_0$ and $\bg_c$ are odd in $y$ and
thus zero in the channel centre.
\begin{subequations}
\label{eq:trigexpand}
\begin{align}
&\bg=\bg_0(y)+\bg_c(y) && z=0:& 
\mbox{Centre of laminar region} \label{eq:g0}\\
&\bg=\bg_0(y)+\bg_s(y) && z=\lambda_z/4=10:& 
\mbox{Laminar-turbulent boundary} \label{eq:g10}\\
&\bg=\bg_0(y)-\bg_c(y) && z=\lambda_z/2=20:& 
\mbox{Centre of turbulent region} \label{eq:g20}\\
&\bg=\bg_0(y)-\bg_s(y) && z=3\lambda_z/4=30:& 
\mbox{Turbulent-laminar boundary} \label{eq:g30}
\end{align}
\end{subequations}

\begin{figure}
\centerline{
\includegraphics[width=14cm]{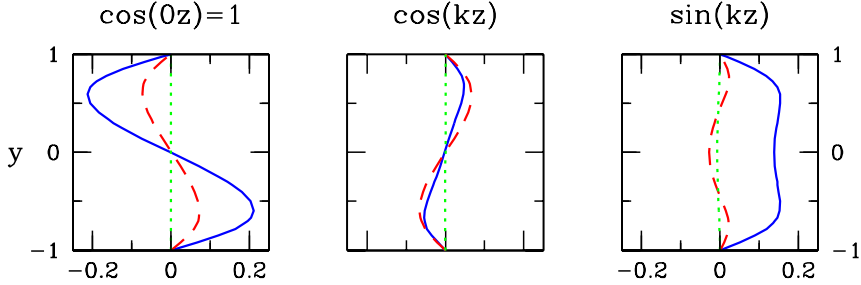}}
\caption{Fourier decomposition of mean velocity. $U$ component (blue, solid),
  $V$ component (green, dotted), $W$ component (red, dashed).  $W_c \approx
  -W_0$, corresponding to the fact that $W$ shows no deviation from the linear
  in the laminar region.
}
\label{fig:fourmean}
\end{figure}

\begin{figure}
 \centerline{
 \includegraphics[width=8cm]{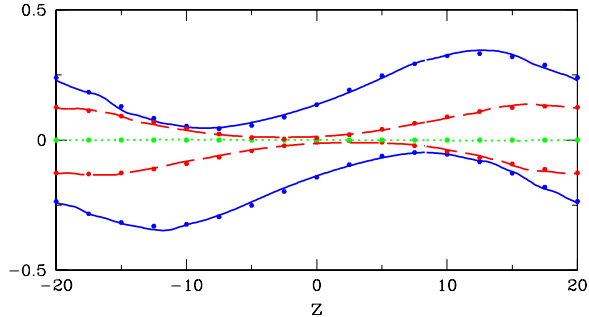}}
 \caption{Mean flow as a function of $z$ at $y=0.725$ (lower curves)
and $y=-0.725$ (upper curves).  $U$ (blue, solid),
   $V$ (green, dotted), $W$ (red, dashed).  Dots show values calculated
from trigonometric formula \eqref{eq:3fcns}.}
 \label{fig:profileZuvw}
 \end{figure}

\begin{figure}
\includegraphics[width=14cm]{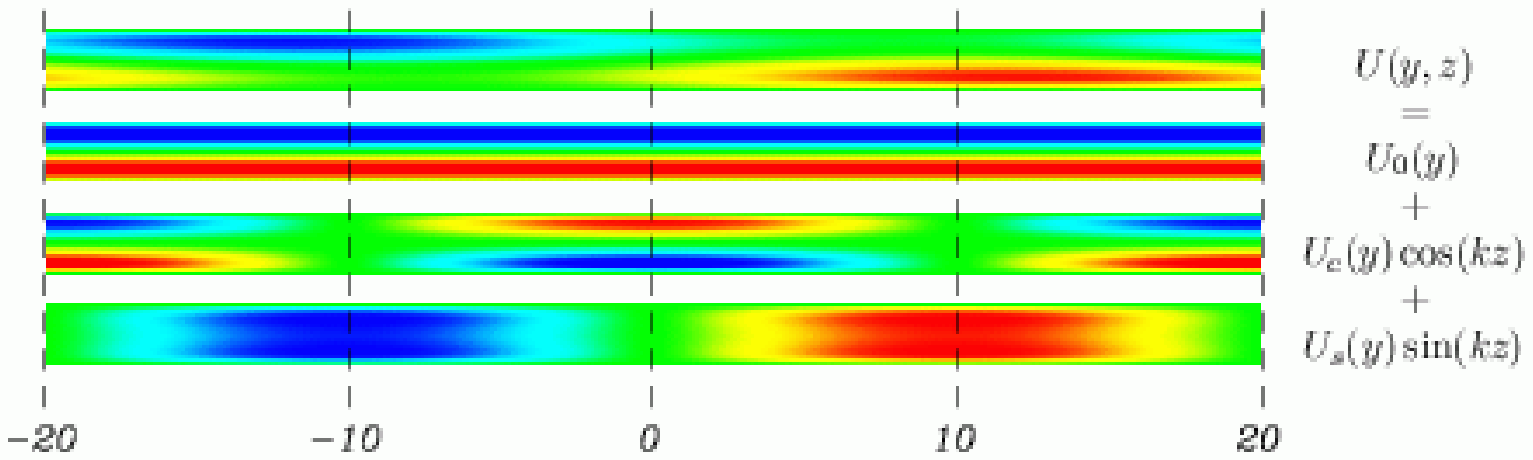}
\caption{Mean velocity $U$ and its trigonometric decomposition.
Because the magnitude of the fields vary, different color
scales are chosen to emphasize qualitative features. 
$U~(-0.2, 0.2)$, $U_0~(-0.2, 0.2)$, $U_c~(-0.05, 0.05)$, $U_s~(-0.154, 0.154)$.}
\label{fig:udecomp}
\includegraphics[width=14cm]{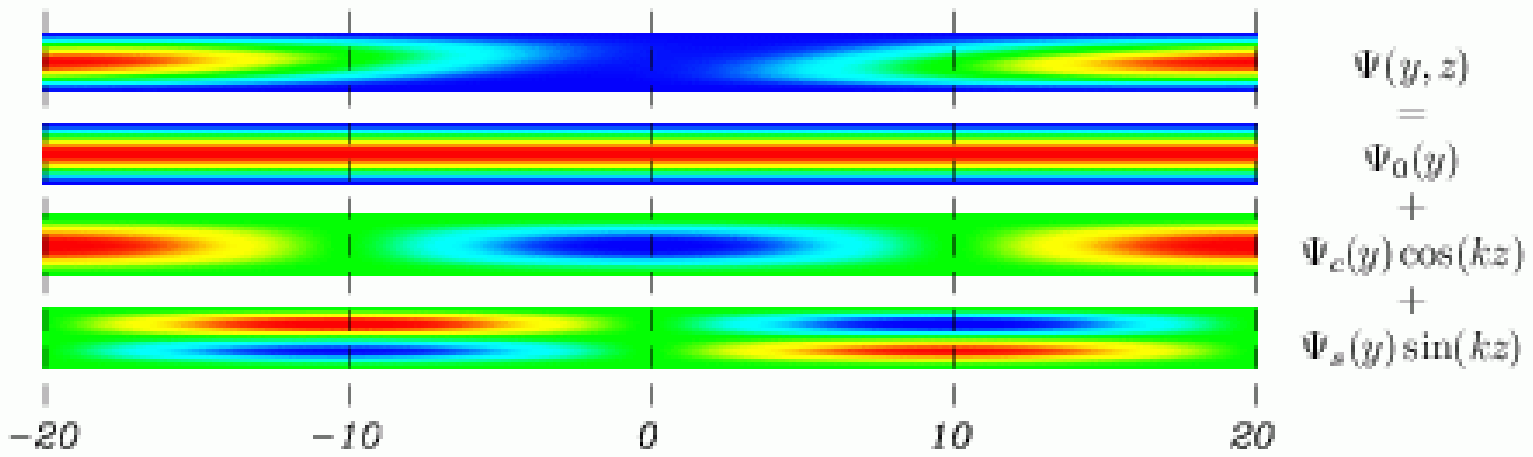}
\caption{Mean streamfunction  $\Psi(y,z)$ for deviation of in-plane flow 
from linear Couette flow and its trigonometric decomposition.
Color scale is
$\Psi~(0, 0.09)$, $\Psi_0~(0, 0.046)$, $\Psi_c~(-0.042, 0.042)$, 
$\Psi_s~(-0.008, 0.008)$.}
\label{fig:psidecomp}
\includegraphics[width=14cm]{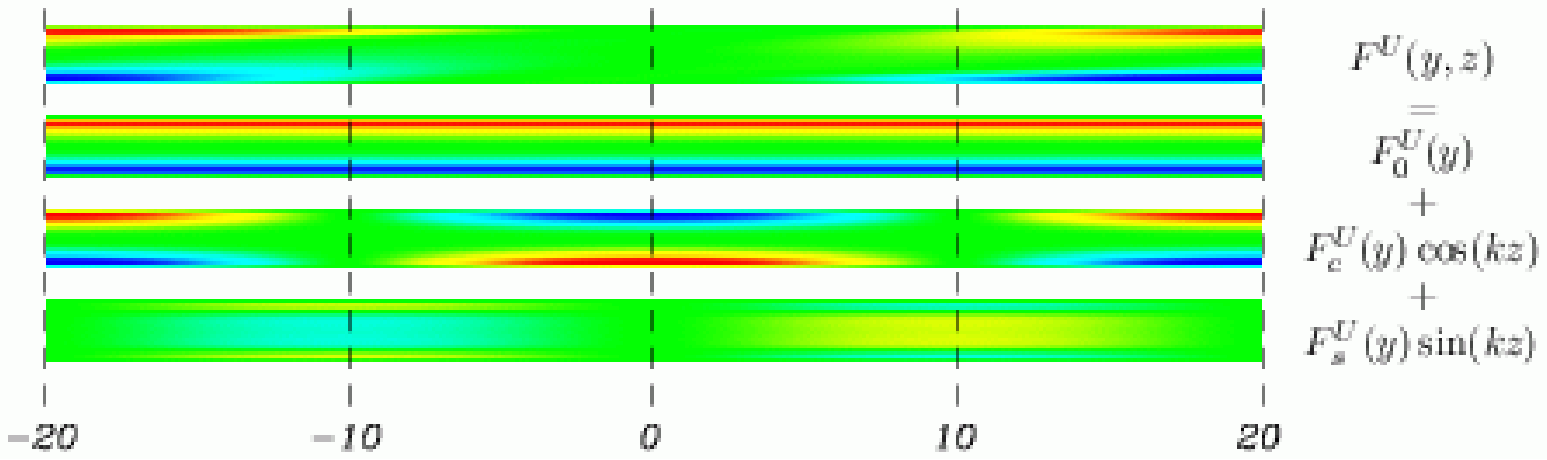}
\caption{Reynolds-stress force $\forcesca^U$ and its trigonometric decomposition.
Color scale is $\forcesca^U~(-0.017,0.017)$, $\forcesca^U_0 (-0.0085,0.0085)$, 
$\forcesca^U_c (-0.0085,0.0085)$, $\forcesca^U_s (-0.0085,0.0085)$.}
\label{fig:Reyudecomp}
\end{figure}

Figure \ref{fig:fourmean} shows the three trigonometric components, each a
function of $y$, obtained by Fourier transforming $U$, $V$, and $W$.  Figure
\ref{fig:profileZuvw} shows $U$, $V$ and $W$ as functions of $z$ at locations
in the upper and lower channel and compares them with the values obtained from
the trigonometric formula \eqref{eq:3fcns} using the functions shown in figure
\ref{fig:fourmean}.
Figures \ref{fig:udecomp}, \ref{fig:psidecomp} and \ref{fig:Reyudecomp}
depict $U(y,z)$, $\Psi(y,z)$ and $\forcesca^U(y,z)$ with their trigonometric
decompositions. Each of these figures uses only the three scalar functions of
$y$, figure \ref{fig:fourmean}, to reproduce the corresponding two-dimensional
field.  As shown by equation~\eqref{eq:definepsi}, the streamfunction $\Psi$
of a centrosymmetric field has symmetry opposite to that of the velocity
components, \ie\ it is composed of functions of the same parity in $y$ and $z$.

\begin{figure}
\centerline{
\includegraphics[width=14cm]{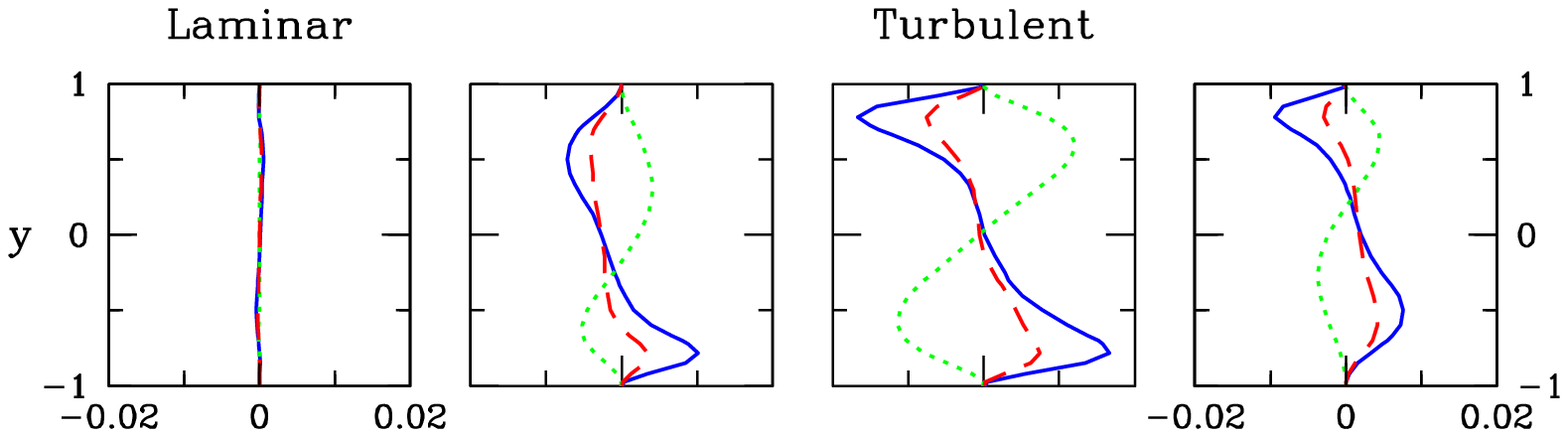}}
\caption{Reynolds-stress force $\forcevec=-\la(\bufluc\cdot\nabla)\bufluc\ra$
as a function of $y$.  Curves show $\forcesca^U$ (blue, solid),
$\forcesca^V$ (green, dotted) and $\forcesca^W$ (red, dashed).}
\label{fig:Reynolds}
\end{figure}

\begin{figure}
\centerline{
\includegraphics[width=14cm]{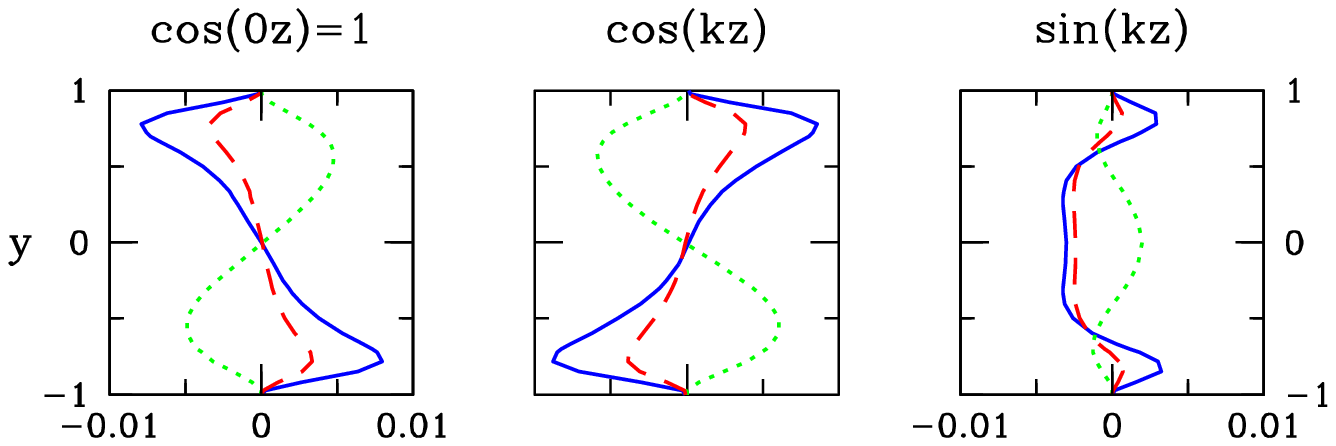}}
\caption{Fourier decomposition of Reynolds-stress force.  $\forcesca^U$ components
  (blue, solid), $\forcesca^V$ components (green, dotted), $\forcesca^W$ components (red,
  dashed).  $\forcevec_c\approx-\forcevec_0$, as required for the vanishing of $\forcevec$ at
  $z=0$.}
\label{fig:fourrey}
\end{figure}

Figures~\ref{fig:Reynolds} and \ref{fig:fourrey} show the three
Reynolds-stress forces and their Fourier decompositions.
Each component obeys $\forcevec_c\approx-\forcevec_0$, a necessary condition for
$\forcevec$ to vanish at the centre of the laminar region, as shown by
equation \eqref{eq:g0} and also illustrated in figure~\ref{fig:Reyudecomp}.
More precisely, 
\begin{equation}
\forcesca^U_c = -1.09 \forcesca^U_0, ~~~~~ \forcesca^V_c = -1.22 \forcesca^V_0, ~~~~~ \forcesca^W_c = -1.16 \forcesca^W_0
\label{eq:falmostzero}
\end{equation}
In addition,
\begin{equation}
\forcevec\approx -\pd_y \la \bufluc\vfluc\ra.  
\end{equation}
as is typical for turbulent channel flows; see, \eg\, \cite{Pope}.
\comment{This may not be the best place for this. Perhaps should
merge with other $\forcevec$ discussion.}

\subsection{Force balance for $U$}
\label{sec:ForceU}

\begin{figure}
\centerline{\includegraphics[width=17cm]{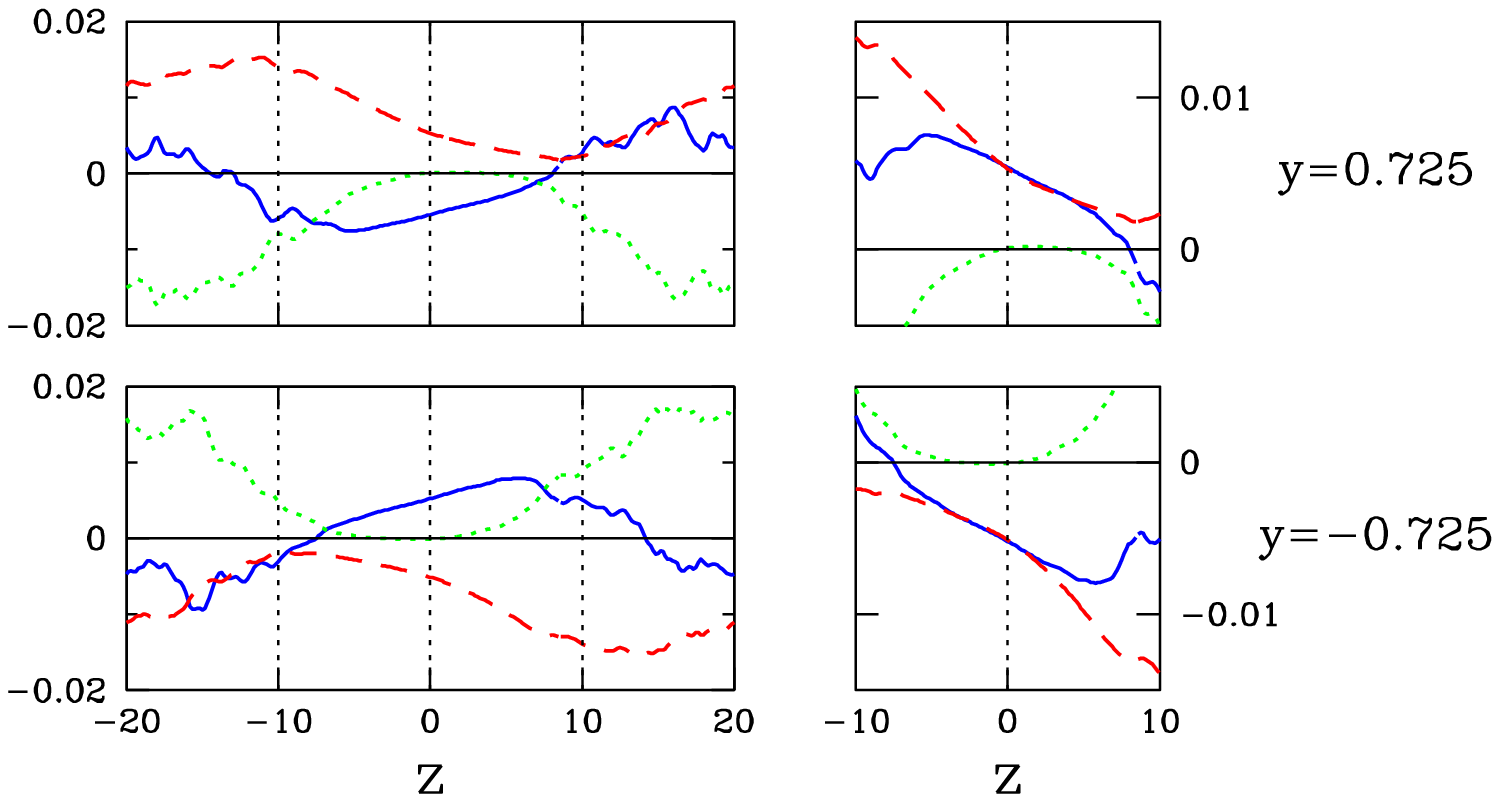}}
\caption{Mean forces in $U$ direction as a function of $z$ at $y=\pm 0.725$
  for turbulent-laminar pattern at $Re=350$.  Advective $-(\bU\cdot\nabla) U$
  (blue, solid), viscous $\lap U$ (red, dashed), and turbulent
  $-\la(\bufluc\cdot\nabla) \ufluc\ra$ (green, dotted) forces.  In the
  laminar region ($z\approx 0$), the Reynolds-stress force vanishes and the
  viscous and advective forces are equal and opposite to one another.  In
  figures on right, enlarged around the laminar region, $\lap U$ and
  $+(\bU\cdot\nabla) U$ are shown to emphasize equality between viscous and
  advective forces.}
\label{fig:balance}
\end{figure}

\begin{figure}
\centerline{
\includegraphics[width=14cm]{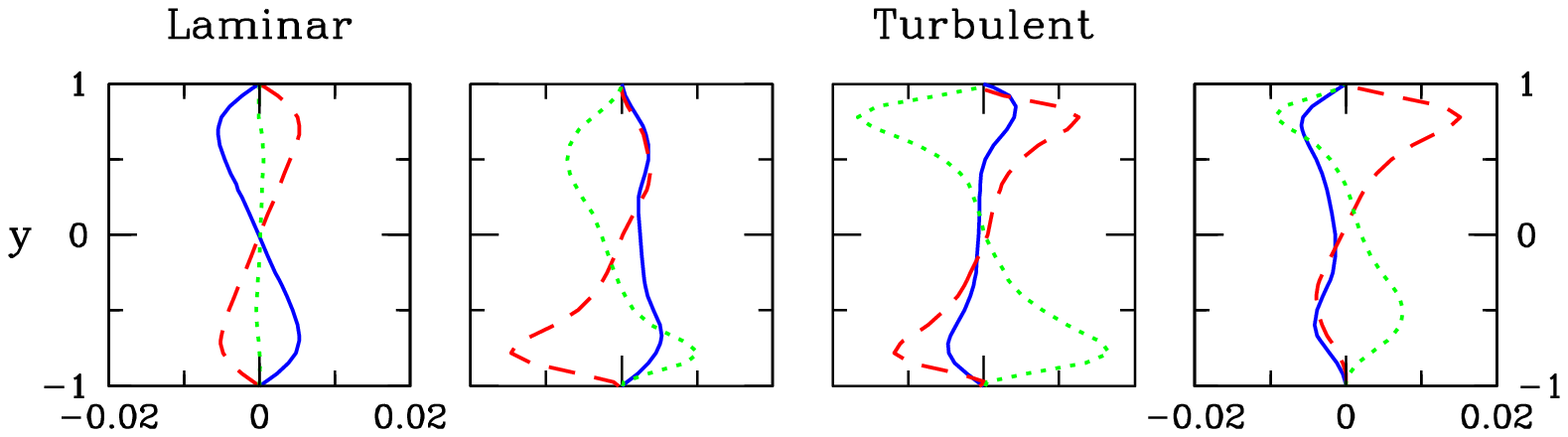}}
\caption{Balance of forces in the $U$ direction.  Curves show advective force
  (blue, solid), viscous force (red, dashed) and Reynolds-stress force (green,
  dotted).  In the laminar region, the Reynolds-stress force is negligible
  and the advective and viscous forces counter-balance one another.}
\label{fig:Forceu}
\centerline{
\includegraphics[width=14cm]{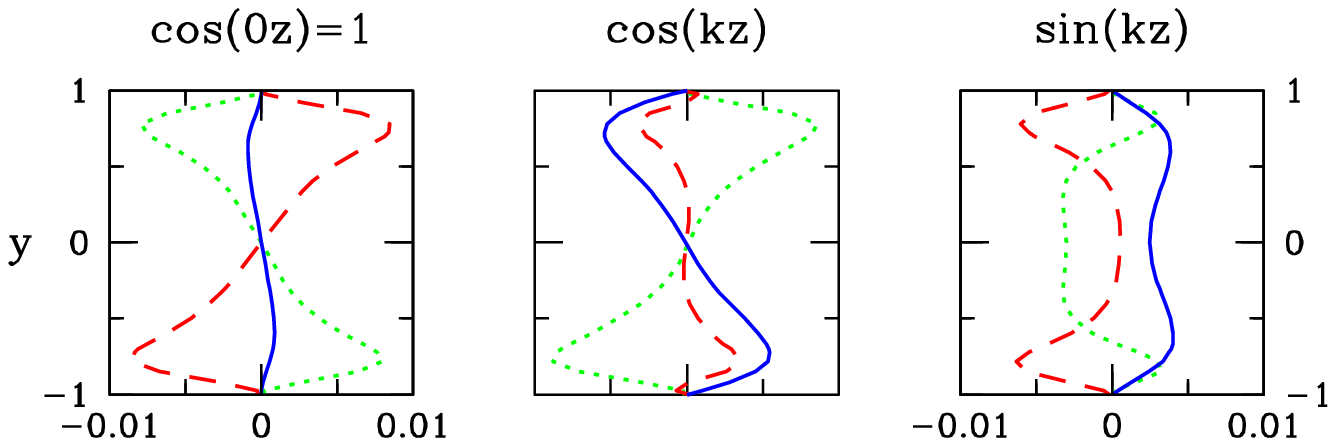}}
\caption{Balance of forces in the $U$ direction, decomposed into modes.
  Curves show advective (blue, solid), viscous (red, dashed) and 
  Reynolds-stress (green, dotted) forces.  
Mode 1: Reynolds-stress and viscous forces approximately 
counterbalance each other. $\cos(kz)$: advection is larger
than viscous force, which is especially small in the bulk.
}
\label{fig:FourForceu}
\centerline{
\includegraphics[width=14cm]{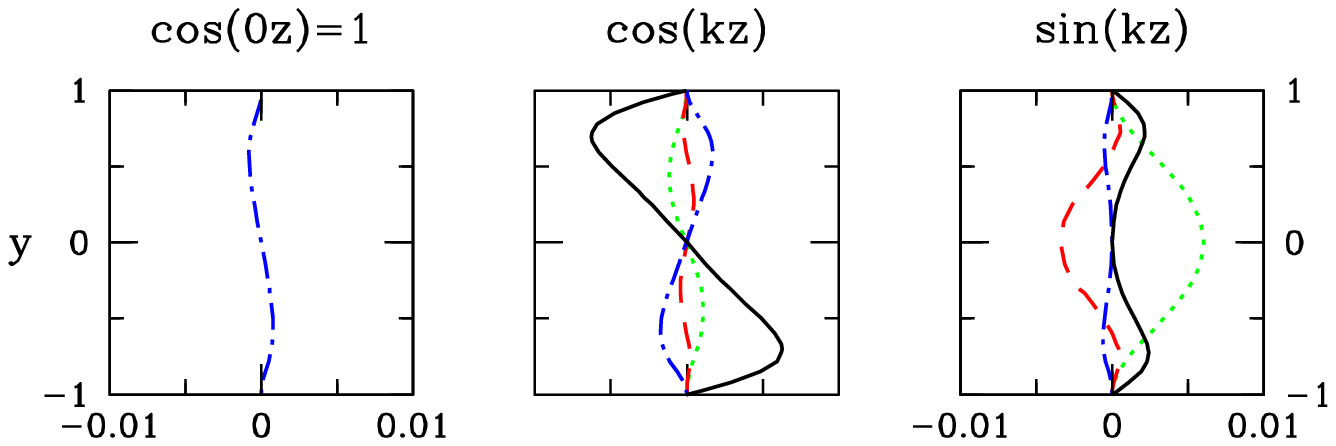}}
\caption{Advective terms in the $U$ direction, decomposed into modes.  
  Curves show $-\beta y \pd_z U$ (black, solid) $-W\pd_z U$ (blue, dash-dot), 
  $-\alpha V$ (green, dotted), $-V\pd_y U$ (red, dashed).
  The 1 mode is generated by the product $-k W_c U_s/2$; 
  a second harmonic of the same small size is also generated.  
  The $\cos(kz)$ mode is dominated by $-\beta\,y\,k\,U_s$.  
  The $\sin(kz)$ term is dominated by $\beta\,y\,k\,U_c$ near the boundaries
  and by $-V_s\pd_y (U_0+\alpha y)$ in the bulk.}
\label{fig:FourNonu}\end{figure}

We now turn to understanding the balance of forces responsible for maintaining
the mean flow profiles. We focus primarily on $U$, both because it is the
component of largest amplitude and also because it appears only in equation
\eqref{eq:NSavgU}: $U$ is subject to Reynolds-stress and viscous forces, and
is advected by $(V,W)$ but is not self-advected.  
We begin by showing the balance of forces in the $U$ direction as a function
of $z$ at locations in the upper and lower channel in
figure~\ref{fig:balance}.  One can again see the centrosymmetry of each of the
forces, \ie\ invariance under the combined operations of reflection in $y$ and
$z$ and change of sign. The Reynolds-stress force disappears at the center of
the laminar region and the advective and viscous forces exactly
counterbalance, as emphasized in the figures on the right.  Figure
\ref{fig:Forceu} shows another view of this balance, displaying the forces as
a function of $y$ at four locations in $z$.
As previously stated, $\lap U$ is dominated by $\pd^2_y U$ and $\forcesca^U$
by $-\pd_y \la \ufluc\vfluc\ra$.
In figure~\ref{fig:FourForceu}, we show the
Fourier-space analogue of figure~\ref{fig:Forceu}.

We now turn to the more complex advective forces, whose Fourier decompositions
are shown in figure \ref{fig:FourNonu}.  The $\cos(0z)$ component of the
advective force is small but non-zero.  Because this term results from the
product of trigonometric functions, it also provides a measure of the
generation of higher harmonics, a point which we will explore further in
section \ref{sec:Model}. The advective $\cos(kz)$ term is well approximated by
the contribution from advection by $w^L=\beta y$.  The advective $\sin(kz)$
term is dominated near the walls by advection by $w^L$, but in the bulk by
advection by $V$.  
Properties of the $\cos(kz)$ and $\sin(kz)$ modes echo
their physical space counterparts: the advective term is well approximated by
advection by $w^L$ in the laminar region, as was shown in figure
\ref{fig:forcepreface}, while the advective forces in the laminar-turbulent
boundaries combine advection by $w^L$ near the walls and by $V$ in the bulk.

\begin{figure}
\centerline{
\includegraphics[width=13cm]{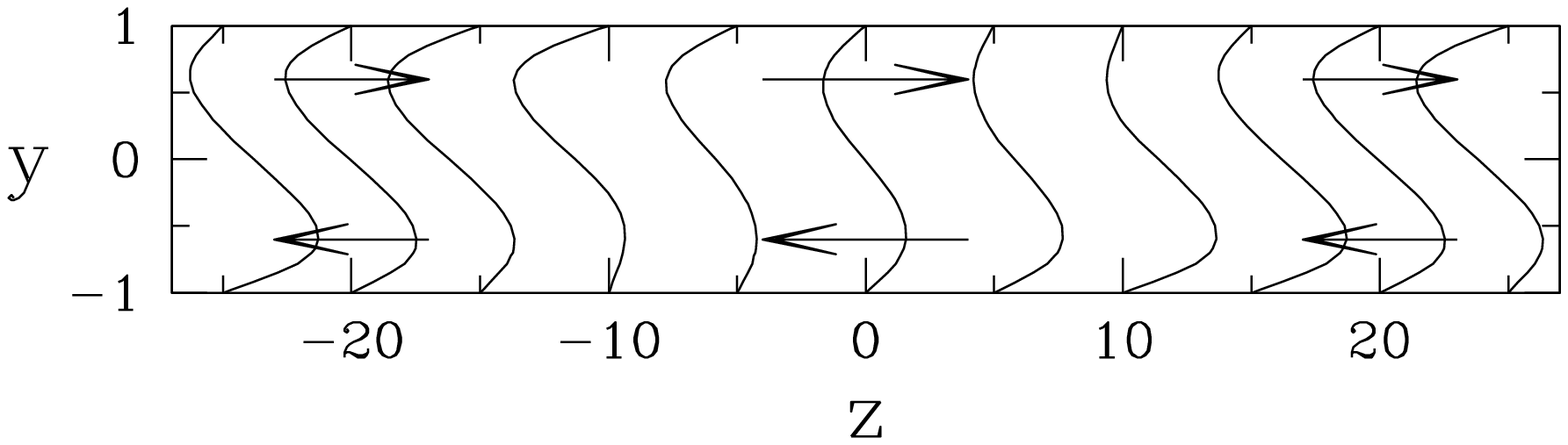}}
\caption{Schematic depiction of the dynamics of $U$ near the centres of the
  laminar and of the turbulent regions. The cross-channel direction is
  exaggerated. Shown are $U$ (profiles) and $W+\beta y$ (arrows). Viscous
  diffusion tends to diminish both peaks of the profile.  In laminar region
  surrounding $z=0$, the peaks in the upper half-channel increase in amplitude
  with decreasing $z$; advection towards positive $z$ (upper arrow)
  replenishes these peaks, maintaining $U$.  Conversely, the peaks in the
  lower half-channel increase with $z$; advection towards negative $z$ (lower
  arrow) replenishes these peaks.  That is, the sign of $-(W+\beta y)\,\pd_z U$
  is opposite to that of $\pd^2_y U$ in both the upper and lower parts of the
  laminar region.  In the turbulent region around $z=\pm 20$, the size of the
  upper (lower) peak decreases to the left (right) and so advection, like
  viscous diffusion, acts instead to diminish the peaks. $U$ is maintained by
  the Reynolds-stress force, which counterbalances both.  The effect is to
  modulate the amplitude of the $U$ profiles periodically in $z$.}
\label{fig:schematic_cos}
\end{figure}

\begin{figure}
\centerline{
\includegraphics[width=13cm]{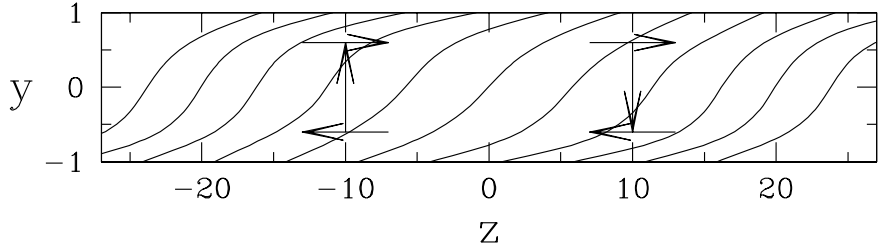}}
\caption{Schematic depiction of the dynamics of $U$ near the turbulent-laminar
  boundaries. The cross-channel direction is exaggerated. Shown are $U+\alpha
  y$ (profiles), and $(V, W+\beta y)$ (arrows).  Near the upper and lower
  walls, the $U+\alpha y$ profiles are advected towards increasing/decreasing
  $z$ by $W + \beta y$.  In the bulk, advection by $V$ is significant.  At
  $z\approx 10$, $V$ advects downwards the right-moving fluid in the upper
  portion of the channel. At $z\approx -10$, $V$ advects upwards the
  left-moving fluid in the lower portion of the channel.  The effect is to
  tilt the $U=0$ boundary periodically in $z$.}
\label{fig:schematic_sin}
\end{figure}

We illustrate these conclusions via schematic visualizations of the dynamics
of $U$.  Figure \ref{fig:schematic_cos} illustrates the dynamics in the
laminar and turbulent regions. The dynamics in the laminar region are
essentially described by the simple balance between viscous diffusion of $U$
profiles and advection by linear Couette flow in $z$, given by equation
\eqref{eq:firstapprox}.  Viscous diffusion tends to reduce curvature, but the
profiles have greater curvature upstream (to the left for the upper channel,
to the right for the lower channel).  Hence advection replenishes the
curvature damped by viscosity.  However, this trend towards greater curvature
upstream cannot continue indefinitely, since the pattern is periodic in $z$.
Hence eventually a maximum is reached (at a turbulent-laminar boundary),
beyond which the curvature decreases upstream.  Thus, in the turbulent region,
advection and diffusion act together to decrease curvature and must both be
counter-balanced by turbulent forcing.  These features are essentially
described by the $\cos(0z)=1$ and $\cos(kz)$ modes.  Figure
\ref{fig:schematic_sin} illustrates the dynamics in the turbulent-laminar
boundaries. These dynamics include advection by $V$ in the bulk, leading to
the $U>0$ ($U<0$) patch in the lower right (upper left) of figure
\ref{fig:udecomp} and are described by the $\sin(kz)$ mode.  \comment{Make
sure this is correct.}

\subsection{Force balance for $W$ and $V$}
\label{sec:ForceWV}

\begin{figure}
\centerline{
\includegraphics[width=14cm]{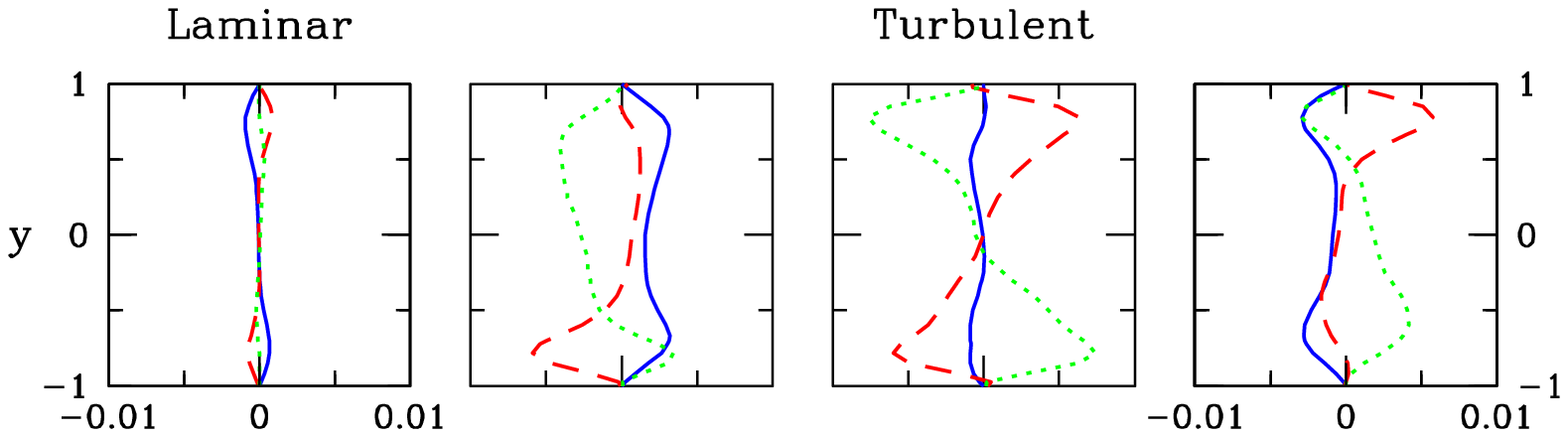}}
\caption{Balance of forces in the $W$ direction.  Curves show advective term
  (blue, solid), viscous force (red, dashed) and Reynolds-stress force (green,
  dotted).  In the laminar region, $W \approx 0$ and each of the forces is
  negligible. In the turbulent region, the viscous and Reynolds-stress forces
  counter-balance one another. In the laminar-turbulent boundaries, the
  advective, viscous and Reynolds-stress forces all play a role.}
\label{fig:Forcew}
\comment{dP/dz is not zero}
\centerline{
\includegraphics[width=14cm]{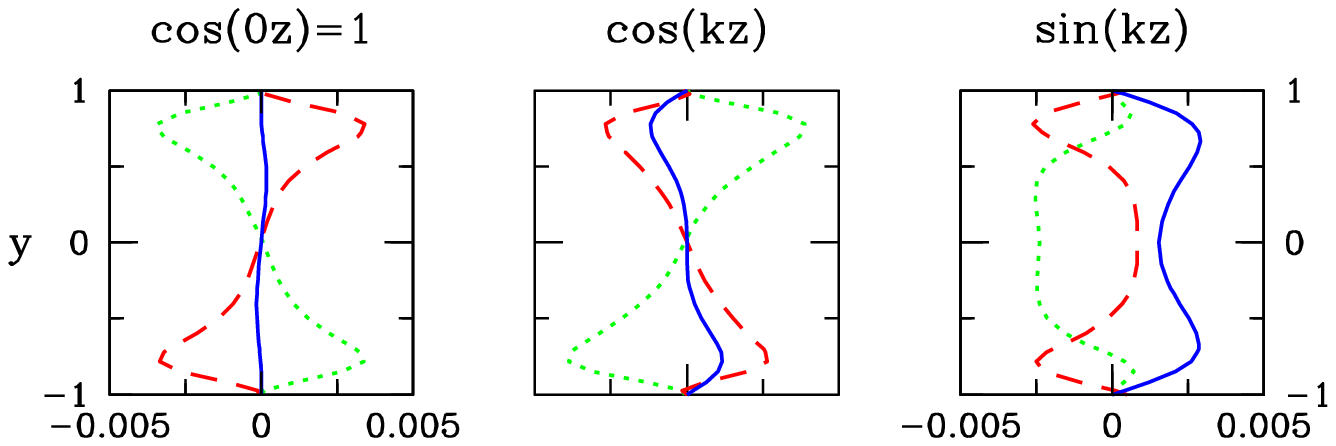}}
\caption{Balance of forces in the $W$ direction, decomposed into modes.
  Curves show advective term (blue, solid), viscous term (red, dashed) and
  turbulent forcing term (green, dotted). 
}
\label{fig:FourForcew}
\centerline{
\includegraphics[width=14cm]{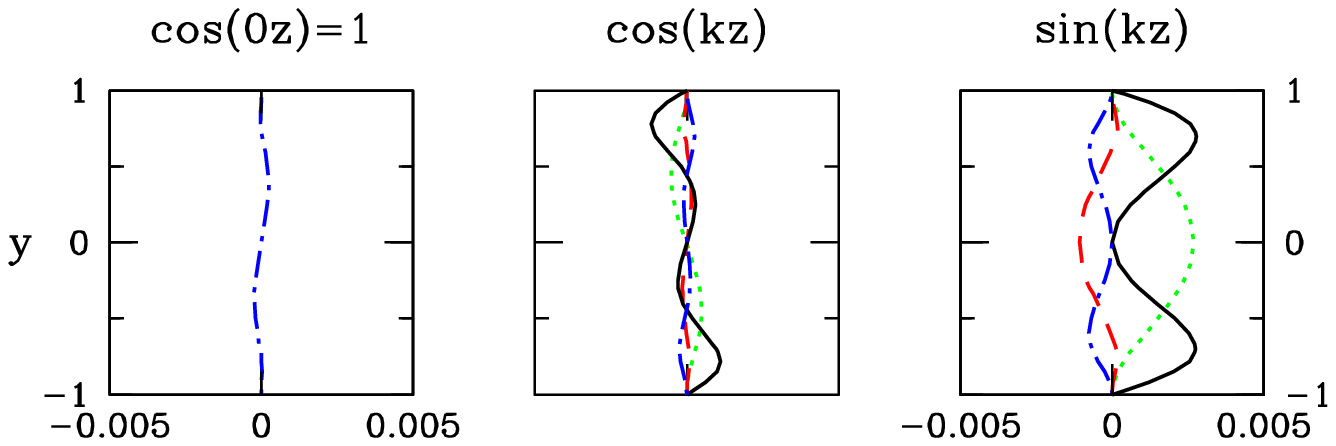}}
\caption{Advective terms in the $W$ direction, decomposed into modes.  
  Curves show $-\beta y \pd_z W$ (black, solid) $-W\pd_z W$ (blue, dash-dot), 
  $-\beta V$ (green, dotted), $-V\pd_y W$ (red, dashed).
  The $\cos(kz)$ mode is dominated by $-\beta\,y\,k\,W_s$.  
  The $\sin(kz)$ term is dominated by $\beta\,y\,k\,W_c$ near the boundaries
  and by $-V_s\pd_y (W_0+\beta y)$ in the bulk.}
\label{fig:FourNonw}
\end{figure}
Figure \ref{fig:Forcew} shows the balance of forces in the $W$ direction and
figure~\ref{fig:FourForcew} its analogue in Fourier space.  This balance
resembles that in the $U$ direction shown in figures \ref{fig:Forceu} and
\ref{fig:FourForceu}.  In physical space (compare the leftmost panels of
figures \ref{fig:Forcew} and \ref{fig:Forceu}), the main difference is that
the advective and viscous forces are both small in the laminar region, in
keeping with the fact that $W\approx 0$.  The pressure gradient $\pd_z P$ is
far smaller than the other forces throughout (see below).  In Fourier space
(compare the middle panels of figures \ref{fig:FourForcew} and
\ref{fig:FourForceu}), the main difference with the $U$ balance is that the
relative importance of the viscous and advective forces in the $\cos(kz)$
balance is reversed from that in the case of $U$: for $W$, the viscous
component is larger than the advective component, which is especially small in
the bulk.  The decomposition of the advective terms (figure
\ref{fig:FourNonw}) shows that, as is the case for $U$, the advective
$\cos(kz)$ term is well approximated by the contribution from advection by
$w^L=\beta y$, whereas all four advective components contribute to the
$\sin(kz)$ term.

\begin{figure}
\centerline{
\includegraphics[width=14cm]{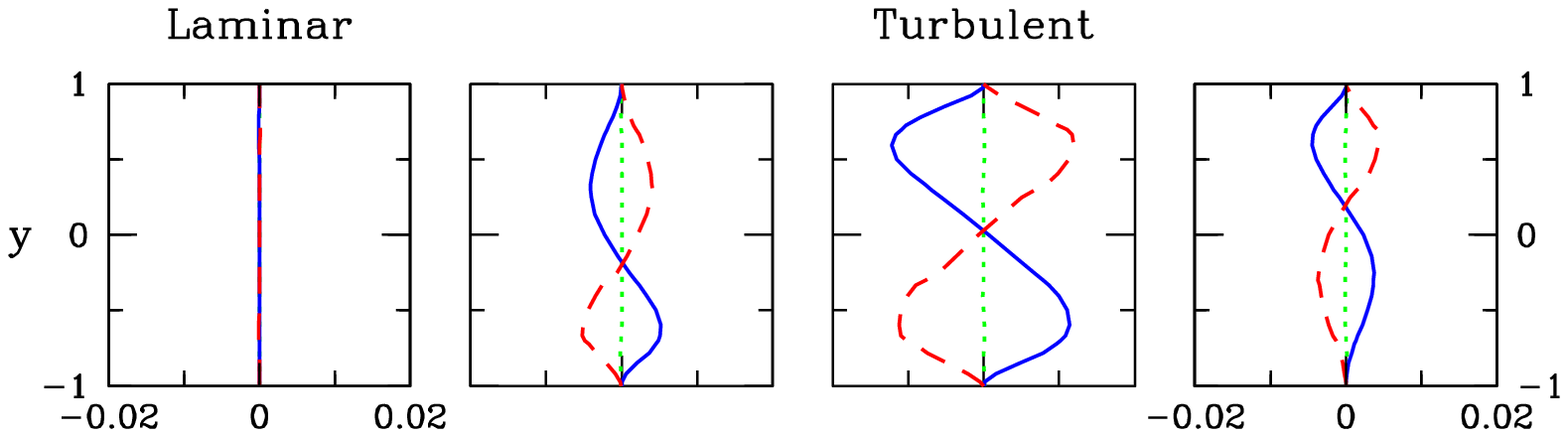}}
\caption{Forces in the $V$ direction.  
  Reynolds-stress force $\forcesca^V$(red, dashed) is
  counterbalanced by pressure gradient $-\pd_y P$ (blue, solid). Both are zero
  in the laminar region. Advective and viscous forces (green, dotted) are
  negligible throughout.}
\label{fig:forcev}
\end{figure}

The balance of forces in the $V$ direction is entirely different.
The dominant balance in this equation is:
\begin{equation}
0 = -\pd_y P + \forcesca^V
\label{eq:Vdombal}\end{equation}
as shown in figure \ref{fig:forcev}.
This is typical for turbulent channel flows; see, \eg\, \cite[]{Pope}.  
This balance between the mean pressure
gradient $P$ and the Reynolds-stress force $\forcesca^V$ does not constrain or provide
information about any of the velocity components.  Since
\begin{equation}
\forcesca^V=-\divv \la \bufluc\vfluc\ra\approx -\pd_y\la\vfluc^2\ra
\end{equation}
\comment{comment about arbitrary function of z being constant because of 
w balance equation}
we in fact have
\begin{equation}
P\approx-\la\vfluc^2\ra
\label{eq:v2fluc}\end{equation}
up to a small $z$-dependent correction.  Figure~\ref{fig:upsikep} shows the
pressure field $P$ calculated from \eqref{eq:v2fluc} and suggests that its $y$
dependence can be approximated by the functional form $\cos(\pi y/2)$.  This
leads to an estimate of the relative importance of the pressure gradients in
the $y$ and $z$ directions:
\begin{equation}
O\left(\frac{\pd_y P}{\pd_z P}\right)= \frac{\pi/2}{k} \approx 10
\label{eq:dyPtodzP}
\end{equation}
while our data shows
\begin{equation}
\frac{(\pd_y P)_{\rm max}}{(\pd_z P)_{\rm max}} = \frac{0.012}{0.0017} = 7.05
\end{equation} 
The same estimate applies to the relative magnitudes of $V$ and $W$, using the
streamfunction shown in figure~\ref{fig:upsikep}:
\begin{equation}
O\left(\frac{W}{V}\right) = 
O\left(\frac{\pd_y \Psi}{\pd_z \Psi}\right) = \frac{\pi/2}{2\pi/40} = 10
\label{eq:WtoV}\end{equation}
while the actual ratio of maximum values is
\begin{equation}
\frac{W_{\rm max}}{V_{\rm max}} = \frac{0.15}{0.013} = 11.
\end{equation} 

\comment{Need a reference for this behavior in ordinary channel flow.}

\subsection{Model equations}
\label{sec:Model}

We now derive a system of ordinary differential equations by substituting the
trigonometric form \eqref{eq:3fcns} into the Reynolds-averaged Navier-Stokes
equations \eqref{eq:RANS}.  The drawback in this procedure is the usual one,
namely that this form is not preserved by multiplication.  However, Table
\ref{tab:fourier} shows that higher harmonics contribute very little to $\bU$.

We expand the advective term as:
\begin{subequations}
\label{eq:nonnon}
\begin{eqnarray}
((\bU+\bu^L)\cdot\grad)&&(\bU+\bu^L) 
= (V\:\pd_y + (W+\beta y)\:\pd_z) (\bU +\alpha y \ex+\beta y \ez)\nonumber\\
&=& (V_c\cos(kz)+V_s\sin(kz))
(\bU\p_0+\bU\p_c\cos(kz) + \bU\p_s\sin(kz) +\alpha \ex +\beta \ez)\nonumber\\
&+&(W_0+\beta y +W_c\cos(kz)+W_s\sin(kz))
(-k\:\bU_c\sin(kz)+k\:\bU_s\cos(kz))\nonumber\\
&=&\half\left(V_c\:\bU\p_c+V_s\:\bU\p_s+k\:(W_c\:\bU_s-W_s\:\bU_c)\right)
\label{eq:non0}\\
&+&(V_c (\bU\p_0+\alpha\ex+\beta\ez)  +k\:(W_0+\beta y)\bU_s)\cos(kz) 
\label{eq:nonc1}\\
&+&(V_s (\bU\p_0 +\alpha\ex + \beta\ez) -k\:(W_0+\beta y)\bU_c)\sin(kz)
\label{eq:nons1}\\
&+&\half\left(V_c\:\bU\p_c-V_s\:\bU\p_s+k\:(W_c\:\bU_s+W_s\:\bU_c)\right)
\cos(2kz),\label{eq:non2}
\end{eqnarray}
\end{subequations}
where primes denote $y$ differentiation.  We neglect the second harmonic term
\eqref{eq:non2}, and will discuss the accuracy of this approximation below.
We now rewrite the $U$ and $W$ components of the averaged momentum equations,
neglecting the $z$-derivatives $\pd^2_z U$, $\pd^2_z W$ and $\pd_z P$, as
justified by equations \eqref{eq:d2ytod2z} and \eqref{eq:dyPtodzP}:
\begin{subequations}
\label{eq:RANSnew}
\begin{eqnarray}
0 &=& -\left(V \pd_y + (W + \beta y)~ \pd_z \right) (U + \alpha y) 
+ \oor \pd^2_y U + \forcesca^U \\
0 &=& -\left(V \pd_y + (W + \beta y)~ \pd_z \right) (W + \beta y) 
+ \oor \pd^2_y W + \forcesca^W 
\end{eqnarray}
\end{subequations}
Substituting \eqref{eq:non0}--\eqref{eq:nons1} in \eqref{eq:RANSnew} and
separating terms in $\cos(0z)=1$, $\cos(kz)$ and $\sin(kz)$, we obtain
\begin{subequations}
\begin{eqnarray}
0 &=& -\half\left[V_c U\p_c + V_s U\p_s + k(W_c U_s  - W_s U_c) \right]
+\oor U\pp_0 + \forcesca^U_0 \label{eq:u0}\\
0 &=& -V_c (U\p_0+\alpha) - k\:(W_0+\beta y) U_s + 
\oor U\pp_c + \forcesca^U_c \label{eq:uc}\\
0 &=& -V_s (U\p_0+\alpha) + k\:(W_0 +\beta y) U_c + 
\oor U\pp_s + \forcesca^U_s \label{eq:us}\\
0 &=& -\half\left[V_c W\p_c+ V_s W\p_s \right]
+\oor W\pp_0 + \forcesca^W_0 \label{eq:w0}\\
0 &=& -V_c (W\p_0+\beta) - k\:(W_0+\beta y) W_s +
\oor W\pp_c + \forcesca^W_c \label{eq:wc}\\
0 &=& -V_s (W\p_0+\beta)  +k\:(W_0 +\beta y) W_c +
\oor W\pp_s + \forcesca^W_s \label{eq:ws}
\end{eqnarray}
\label{eq:UWeqs}\end{subequations}
where the Fourier modes of $V$ and $W$ are related via those of the
streamfunction $\Psi$ of \eqref{eq:definepsi}:
\begin{subequations}
\label{eq:Psicomps}
\begin{align}
&V_0 = 0 & W_0 = \Psi\p_0 \\
&V_c = -k\Psi_s & W_c = \Psi\p_c \\
&V_s = k\Psi_c & W_s = \Psi\p_s
\end{align}
\end{subequations}
and where homogeneous boundary conditions are imposed:
\begin{subeqnarray}
0&=& U_0 = U_c = U_s \;\;\;\;\mbox{at}\;y=\pm 1 \\
0&=& W_0 = W_c = W_s \;\;\;\;\mbox{at}\;y=\pm 1 
\label{eq:UWbcs}\end{subeqnarray}
\comment{Check boundary conditions on Psi. Again, boundary layer
and/or Blasius}

System \eqref{eq:UWeqs} with boundary conditions \eqref{eq:UWbcs} is composed
of six ordinary differential equations coupling the six scalar functions
$U_0,U_c,U_s,\Psi_0,\Psi_c,\Psi_s$ of $y$, with six turbulent forces
$\forcesca^U_0,\forcesca^U_c,\forcesca^U_s,\forcesca^W_0,\forcesca^W_c,\forcesca^W_s$.

\begin{figure}
\centerline{
\includegraphics[width=14cm]{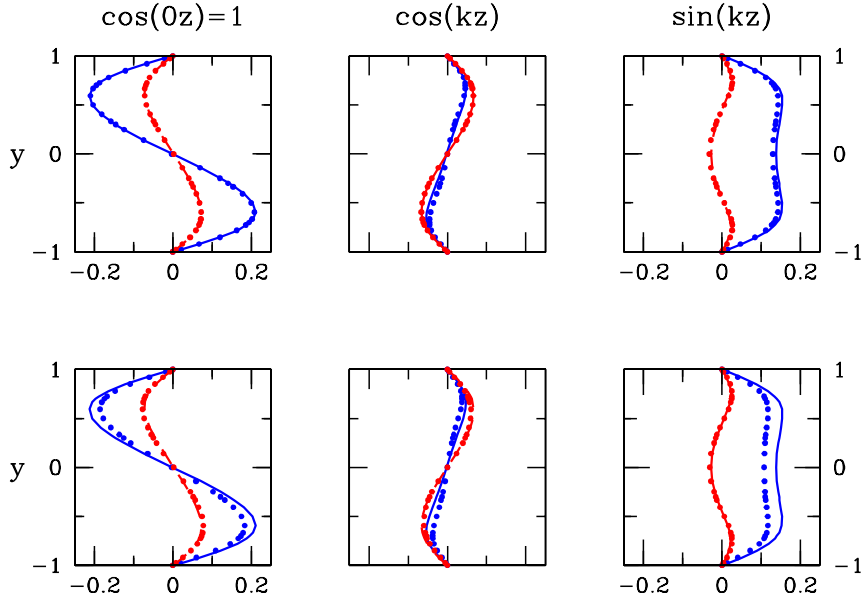}}
\caption{Comparison between mean velocities (in Fourier representation) from
full DNS and ODE models.  
Top row: Curves show $U$ (blue, solid) and $W$ (red, dashed) from DNS.
Dots show solution to the full ODE model \eqref{eq:UWeqs},
essentially indistinguishable from the solid curves. 
Bottom row: Curves again show $U$ (blue, solid) and $W$ (red, dashed) from DNS.
Dots show solution to simplified ODE model \eqref{eq:UWeqssimp}. The
agreement with DNS is very good, though there are small differences
particularly in the $U$ component.}
\label{fig:fourdb}
\end{figure}

We have solved \eqref{eq:UWeqs}--\eqref{eq:UWbcs} numerically, using as inputs
$\forcesca^U$ and $\forcesca^W$ obtained from our full simulations, \ie\ the $\forcevec$ modes shown
in figure \ref{fig:fourrey}.  The resulting solutions are shown in
figure~\ref{fig:fourdb}.  For comparison, we reproduce from figure
\ref{fig:fourmean} the mean velocity fields, in Fourier representation, from
our full simulations (DNS). The ODE solutions are virtually indistinguishable
from the mean fields from DNS.  Only in the sine component of $U$ can the ODE
solutions be distiguished (and only very slightly) from the DNS results.
From the profiles in figure~\ref{fig:fourdb}, the full mean fields could be
constructed as in figures~\ref{fig:udecomp} and \ref{fig:psidecomp}.
Thus, while the ODE model requires input of the Reynolds-stress force terms,
$\forcesca^U$ and $\forcesca^W$, it demonstrates the simplicity of the force balance
responsible for generating the patterned flow when viewed in the Fourier
representation.
Considering higher harmonics would be straightforward, but would serve little
purpose.

We can go in the other direction and attempt to simplify system
\eqref{eq:UWeqs}. The approximate equalities $\forcesca^U_c\approx -\forcesca^U_0$,
$\forcesca^W_c\approx -\forcesca^W_0$ [see equation \eqref{eq:falmostzero}], necessary for
$\forcevec$ to vanish at the center of the laminar region, can be imposed exactly,
reducing the number of turbulent forcing input functions to four.  The terms
arising from the advective forces can be reduced by making approximations
justified from figures \ref{fig:FourNonu} and \ref{fig:FourNonw}.  The
nonlinear terms in \eqref{eq:u0} and \eqref{eq:w0} can be neglected.  The
advective terms in \eqref{eq:uc} and \eqref{eq:wc} can be approximated by
$-k\,\beta \,y \,U_s$ and $-k\,\beta\, y \,W_s$.  Making these approximations, we obtain:

\begin{subequations}
\begin{eqnarray}
0 &=& \oor U\pp_0 + \forcesca^U_0 \label{eq:u0simp}\\
0 &=& - k \,\beta \,y \,U_s + \oor U\pp_c - \forcesca^U_0 \label{eq:ucsimp}\\
0 &=& - V_s (U\p_0+\alpha) + k (W_0+\beta y) U_c 
+ \oor U\pp_s +\forcesca^U_s \label{eq:ussimp}\\
0 &=&  \oor W\pp_0 + \forcesca^W_0\label{eq:w0simp}\\
0 &=& - k \,\beta \,y \,W_s + \oor W\pp_c - \forcesca^W_0\label{eq:wcsimp}\\
0 &=& -V_s (W\p_0 +\beta) + k (W_0+\beta y) W_c 
+ \oor W\pp_s + \forcesca^W_s \label{eq:wssimp}
\end{eqnarray}
\label{eq:UWeqssimp}
\end{subequations}
The solutions to this simplified ODE model are also presented in
figure~\ref{fig:fourdb}.  There is quite good agreement with full DNS results,
thus demonstrating that the dominant force balance is captured by this very
simple system of ODEs. We stress that the only nonlinearities in this model
are in equations~\eqref{eq:ussimp} and \eqref{eq:wssimp}. This reflects the
complexity of the dynamics in the turbulent-laminar boundaries regions (and
the simplicity of the dynamics in the centre of the turbulent and laminar
regions.)

From the simplified ODE model we can obtain the approximate equation satisfied
at the centre of the laminar region by adding equations~(\ref{eq:u0simp}) and
(\ref{eq:ucsimp}):
\begin{equation}
k\,\beta\, y \, U_s= \oor (U_0+U_c)\pp.
\label{eq:lam}
\end{equation}
This is a restatement in terms of Fourier components of the balance described
by equation \eqref{eq:firstapprox} and figure \ref{fig:forcepreface}.

\section{Discussion}

We have presented an analysis of a particular turbulent-laminar pattern
obtained in simulations of large-aspect-ratio plane Couette flow.  We have
focused on a single example so as to understand in quantitative detail the
structure of these unusual flows.
The key findings obtained in our study are as follows. 
First we find that in the (quasi-) laminar flow region the velocity profiles
are not simply those of linear Couette flow. Instead a non-trivial flow is
maintained in the laminar regions by a balance between viscous diffusion and
nonlinear advection.
Next we have considered the symmetries of the flow.  When the pattern forms,
the time-averaged flow breaks the translation symmetry but not centrosymmetry. 
The patterned state is centrosymmetric about the centre of the laminar region
and about the center of the turbulent region.
Next we have considered a spatial Fourier decomposition of the mean flow in
the direction of the pattern wavevector.  From this we find that the lateral
structure of the pattern is almost completely harmonic, \ie\ composed of a
constant and single harmonic.
Thus the pattern description can be reduced to just three cross-channel 
functions for each field, in that
$\bU(x,y,z) \approx \bU_0(y) + \bU_c(y) \cos(kz) + \bU_s(y)\sin(kz)$.
The absence of higher harmonics suggests that the pattern is near 
the threshold, in some sense, of a linear instability of a uniform 
turbulent state.
Such an instability would be governed by a linear equation with 
coefficients which are constant in $z$, whose solutions are 
necessarily trigonometric in $z$.

From our analysis of the turbulent-laminar pattern, in particular its Fourier
decomposition, we derive a model which reproduces the patterned flow.  The
model is derived from the averaged Navier-Stokes equations with the following
assumptions. The crucial assumption, which is strongly supported by our
numerical computations, is that the mean flow can be expressed in terms of
just three horizontal modes. Effectively the generation of higher harmonics
via nonlinear terms in the Navier-Stokes equations is negligible in the mean
flow.
The model is then further simplified because viscous diffusion is
dominated by cross-channel diffusion -- the standard boundary-layer
approximation -- and because pressure variation is negligible along
the pattern wavevector.
We take as input to the model the Reynolds-stress forces measured from
computations.  Assuming that the Reynolds stresses exactly vanish in the
centre of the laminar regions, the number of inputs to the model is just four
cross-channel functions.
The result is a system of six simple ordinary differential equations which
depend on four forcing functions. The model equations accurately reproduce the
mean flow from full direct numerical simulations.

A number of other researchers have attempted to reduce the description of
turbulent or transitional plane Couette flow by various means.  At these low
Reynolds numbers, there is no doubt that fully resolved direct numerical
simulation is feasible and gives accurate results.  The purpose of formulating
a reduced description is therefore to yield understanding.  We now comment on
the differences between the approaches used by other authors and our
reduction.

In parallel with their experiments, Prigent \etal\
(\citeyear{Prigent_PRL,Prigent_PhysD}) considered a pair of coupled
Ginzburg-Landau (GL) equations with additive noise as a model for the
transition from uniform turbulence to turbulent-laminar banded patterns via
noisy (intermittent) patterns.
These equations describe the variation in time and spanwise
coordinate of the amplitudes $A^{\pm}$ of two sets of laminar bands at
opposite tilt angles. These laminar bands modulate the uniform turbulence in
competition with one another.
Each equation separately has one reflection symmetry which corresponds
physically to the centrosymmetry $\cc$ (equation \eqref{eq:Cdef}) of a banded
pattern.
The coupled GL equations possess a second reflection symmetry, corresponding
physically to a spanwise reflection, which takes the amplitude $A^+$ to $A^-$
and vice versa.
By design, this symmetry is not present in our numerical computations.
Prigent \etal\ used their experimental results to fit the parameters in the GL
equation and then compared simulations of the equations with experimental
results.
Steady patterns in the resulting GL equations have only one non-zero
amplitude and this amplitude possesses the reflection symmetry corresponding to
$\cc$.
Hence, the steady patterns in these simulations have exactly the symmetries of
the patterns we have considered.

An important class of models aims at reproducing dynamics of streamwise
vortices and streaks 
in plane Couette turbulence by using a small number of ordinary differential
equations (ODEs).
These equations describe the time-evolution of amplitudes of modes with fixed
spatial dependence.  \cite{Waleffe_97}, guided by the discovery of the
self-sustaining process (SSP) in direct numerical simulations 
\cite[]{Hamilton},
derived a system of eight ODEs, whose variables represent amplitudes of the
key ingredients of the SSP, namely longitudinal vortices, streaks, and streak
waviness.  This model was later also studied and extended by
\cite{Dauchot_Vioujard} and by \cite*{Moehlis_NJP}.

Two other Galerkin projection procedures have been used to derive ODE models.
The most energetic streamwise-independent modes in a principle orthogonal
composition has been used as a basis for a 13-equation model
\cite*[]{Moehlis_PF} exhibiting heteroclinic cycles; when streamwise-dependent
modes are added, the resulting 31-equation model \cite*[]{Moehlis_JFM}
reproduces elements of the SSP cycle.  Eckhardt and co-workers
\cite[]{Schmiegel,Mersmann} have proposed a Fourier space truncation of the
Navier-Stokes equations in all three spatial directions leading to a
19-equation model.  They calculated turbulent lifetimes and saddle-node
bifurcations giving rise to new steady states in this model.

Manneville and co-workers \cite[]{Manneville_Locher,Lagha_Manneville} have
proposed a drastic Galerkin truncation in the cross-channel direction $y$,
retaining one or two trigonometric (for free-slip boundary conditions) or
polynomial (for rigid boundary conditions) basis functions, but fully
resolving both lateral directions.  Simulating the resulting PDEs using a
Fourier basis, they have been able to study phenomena such as the
statistics of lifetimes of turbulent spots in domains with very large lateral
dimensions.

The reduction we have presented differs from the aforementioned studies in
several respects.  Most importantly, we do not describe any time-dependent
behaviour.  We consider here neither turbulent-laminar patterns which are
themselves dynamic (as in Prigent \etal), nor do we consider the
dynamics of streaks and vortices within the turbulence, nor do we consider the
transient dynamics of turbulence.  Instead we focus on the spatially
periodic mean flow of steady turbulent-laminar patterns.  While the turbulent
portions of patterns are dynamic, containing streaks and streamwise vortices,
these are on a fine scale relative to spatial scales of interest here.
Our model description follows directly from an analysis of full numerical
simulations (not from any {\em a priori} assumptions, physical or
phenomenological), that show that all averaged velocity components and forces,
including the Reynolds stress force, are almost exactly trigonometric in the
direction of the pattern wavevector.  It follows directly that the steady
Reynolds-averaged Navier-Stokes equations can be reduced to 6 ODEs for
cross-channel profiles of the Fourier modes.

One of the more significant aspects of this work is the consideration
of the force balance in just the laminar region.  This balance is expressed by
simple equations either in physical space, equation \eqref{eq:firstapprox}, or
in Fourier space, equation \eqref{eq:lam}.
These equations are particularly interesting because they do not contain the
Reynolds stresses, as these are negligable in the laminar region, and hence
their implications can be understood without the need for closure assumptions.

As noted in \S\ref{sec:meanflow}, equation \eqref{eq:firstapprox} implies that
a non-zero tilt angle is necessary to maintain the S-shaped $U$ profile in the
laminar region. If the patterns were not tilted, the flow would necessarily be
laminar Couette flow in the centre of the laminar regions where the turbulence
vanishes.  
We can also derive implications for the relationship 
between Reynolds number, tilt angle and wavelength of the patterns
from equation \eqref{eq:firstapprox}, which we rewrite as:
\begin{equation}
\frac{Re\,\sin\theta}{\lambda} = \frac{(U_0+U_c)\pp}{2\pi \, y \, U_s}
\label{eq:firstapproxrewrite}
\end{equation}
Except where $y\,U_s\approx 0$,
the function on the right-hand-side is indeed approximately
constant across the channel, between about 2.8 and 3.6.
The value of $Re\,\sin\theta/\lambda$ used in our simulations is 
$350 \,\sin(24^\circ)/40=3.56$.

\begin{figure}
\begin{minipage}{6cm}
\centerline{
\includegraphics[width=5cm]{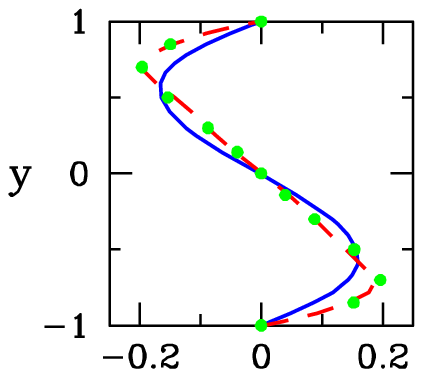}}
\end{minipage}
\begin{minipage}{7cm}
\vspace*{.7cm}
\centerline{
\includegraphics[width=6.2cm]{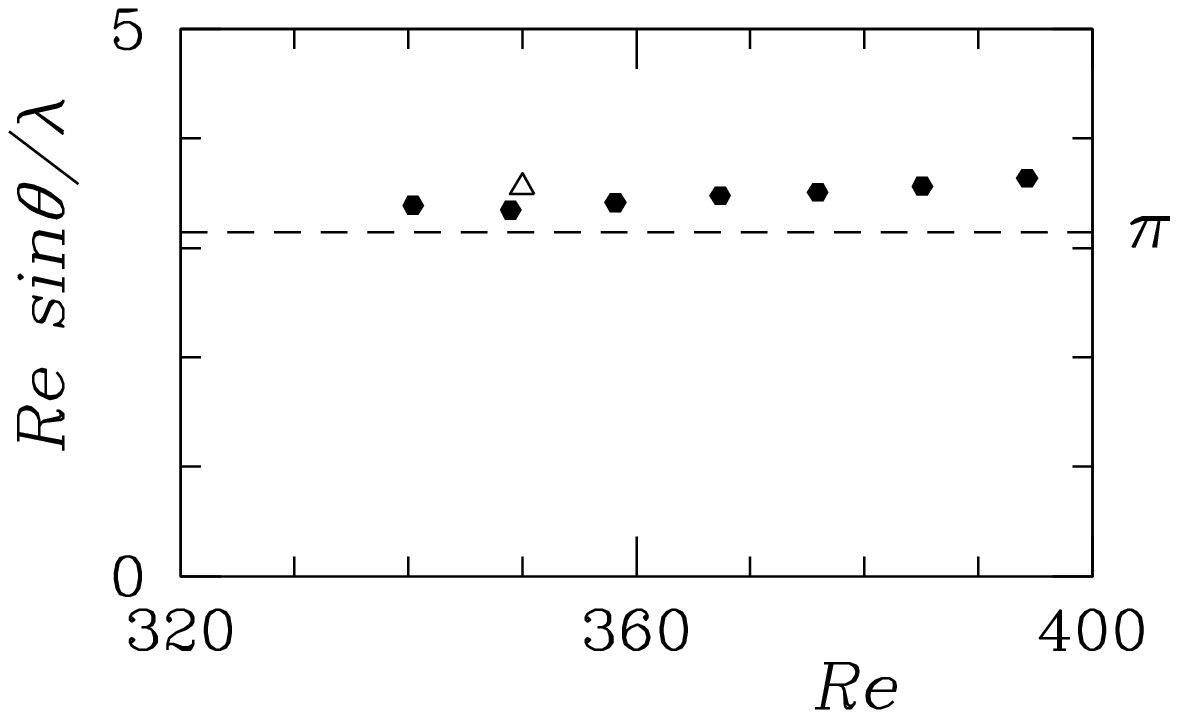}}
\end{minipage}
\caption{Left: Comparison of profiles of $U_0+U_c$ (solid), 
$(U_0+U_c)\pp/(-\pi^2)$ (dashed) and $-2\,y\,U_s$ (points).
The profile of $(U_0+U_c)\pp/(-\pi^2)$ is close to $U_0+U_c$,
in accordance with the approximate functional form $(U_0+U_c)\sim \sin(\pi y)$.
Note that $(U_0+U_c)\pp/(-\pi^2)$ is
almost indistinguishable from $-2\,y\,U_s$, showing that
$(U_0+U_c)\pp/(2\pi \, y \, U_s)$ is near $\pi$ 
over the entire channel.
\newline
Right:
Plot of $\Rey \,\sin\theta/\lambda $ as a function of $\Rey$ for the
experimentally observed patterns of Prigent \etal\ (\citeyear{Prigent_PhysD},
\citeyear{Prigent_IUTAM}). The open triangle shows 
$\Rey \,\sin\theta/\lambda=3.56$
for the case studied numerically in this paper.}
\label{fig:Re_beta_lambda}
\end{figure}

We may obtain a qualitative understanding of this constant as follows;
see figure \ref{fig:Re_beta_lambda} (left).
Observe that in the center of the laminar
region, the functional form of $U = U_0+U_c$ is like $\sin(\pi y)$. Hence its
second $y$ derivative can be approximated by multiplication by $-\pi^2$, or
equivalently $(U_0+U_c)\pp/(-\pi^2) \approx U_0+U_c$. We also find that 
the odd function $-2 \,y \,U_s$ is close to $U_0+U_c$ and
is in fact almost indistinguishable from $(U_0+U_c)\pp/(-\pi^2)$.
This implies that the right-hand-side of equation
\eqref{eq:firstapproxrewrite} is nearly constant across the channel
and equal to $\pi$,
leading to:
\begin{equation}
\frac{\Rey \, \sin\theta}{\lambda} \approx \pi
\label{eq:firstapprox2}
\end{equation} 

\begin{figure}
\centerline{
\includegraphics[width=11cm]{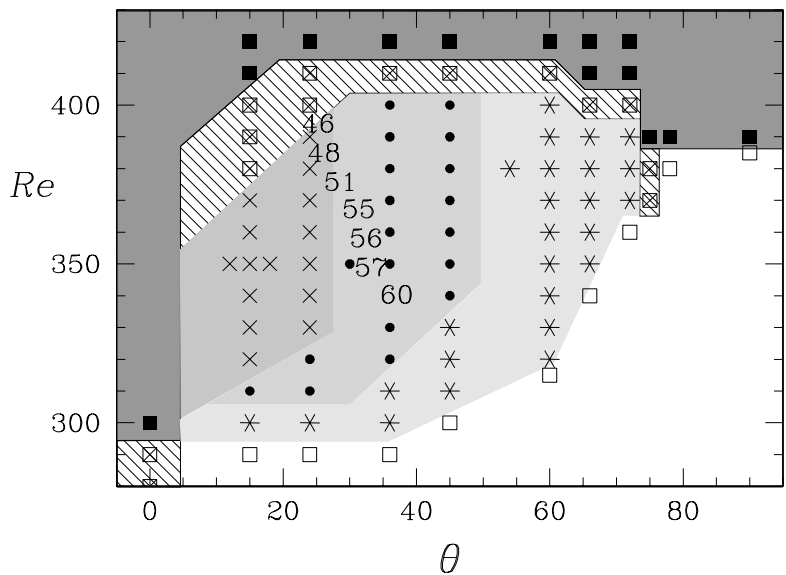}}
\caption{Patterns as a function of Reynolds number $Re$ and $\theta$ in a
computational domain of size $L_x \times L_y \times L_z= (4/\sin\theta)\times
2 \times 120$.  Turbulent-laminar patterns with wavelength
$\lambda=40$ ($\times$), $\lambda=60$ ($\bullet$), $\lambda=120$
($\ast$). Uniform turbulence ($\blacksquare$), intermittent turbulence
($\boxtimes$), laminar Couette flow ($\square$).  Wavelengths in computations
are constrained to be divisors of 120.  Numbers are wavelengths of
experimentally observed patterns of Prigent \etal\ (\citeyear{Prigent_PhysD},
\citeyear{Prigent_IUTAM}).}
\label{fig:angles}
\end{figure}

We believe that equation \eqref{eq:firstapprox2} provides a good first
approximation for the relationship between $\Rey$, $\lambda$, and $\theta$.
Figure \ref{fig:Re_beta_lambda} (right) shows a plot of $\Rey \,
\sin\theta/\lambda$ as a function of $\Rey$ from the experimental data of
Prigent \etal\ (\citeyear{Prigent_PhysD}).  It can be seen that this
combination of quantities is approximately constant with a value near $\pi$.
The range of values of the individual factors $\Rey$, $\theta$, and $\lambda$
can be seen in table \ref{tab:exp_obs}.
In prior studies \cite[]{Barkley_PRL,Barkley_IUTAM}, we have studied a large
range of Reynolds numbers and tilt angles in a domain of length $L_z=120$. In
this domain, the wavelength of a periodic pattern is less constrained, though
it must be a divisor of 120.  Figure~\ref{fig:angles} shows the observed
states as a function of $\Rey$ and $\theta$.  Equation \eqref{eq:firstapprox2}
captures the correct order of magnitude of $\Rey \, \sin\theta/\lambda$;
specifically $1.8 \lesssim \Rey \, \sin\theta/\lambda \lesssim 5$.
Moreover, in figure~\ref{fig:angles} one sees that for fixed $Re$, $\lambda$
increases with increasing $\theta$, as $\eqref{eq:firstapprox2}$ predicts.

Equation \eqref{eq:firstapprox2} does not hold in detail, however.
Most notably, figure~\ref{fig:angles} shows that when $Re$ is decreased
at fixed $\theta$, the wavelength $\lambda$ increases rather than decreases as
one would expect from $\eqref{eq:firstapprox2}$.
We believe that the force balance \eqref{eq:firstapprox} holds for all
patterns which possess a laminar region free of turbulence, but that
the additional approximations made in deriving the simple relationship
\eqref{eq:firstapprox2} do not hold over the full range of
conditions considered in figure~\ref{fig:angles}.
In particular, the right-hand-side of \eqref{eq:firstapproxrewrite} 
depends implicitly on $Re$, $\theta$, and $\lambda$ via the 
dependence of $U_0$, $U_c$, and $U_s$ on these quantities.
The approximate functional relationships between
$U_0$, $U_c$ and $U_s$ that we have observed in our simulations and 
on which we have relied in deriving \eqref{eq:firstapprox2} 
may not hold for other parameter values.
Finer adjustments must come from another mechanism.

The main issue not addressed in our study is closure.  We have not attempted
to relate the forcing of the mean flow due to Reynolds stresses back to the
mean flow itself.
In the future we will report on studies employing closure models.

\acknowledgments

We thank F. Daviaud, O. Dauchot, P. Le Gal, P. Manneville and A. Prigent for
helpful comments.
The simulations analyzed in this work were performed on the
IBM Power 4 of the IDRIS-CNRS supercomputing center as part of project 1119.
This work was supported in part by a CNRS-Royal Society grant.

\oneappendix
\section{Turbulent-laminar bands in other shear flows}

\renewcommand{\arraystretch}{1.5}
\begin{table}
\begin{center}
\begin{tabular}{|c|cc|cc|cc|c|}\cline{1-8}
& 
\multicolumn{2}{|c|}{PC} & \multicolumn{2}{|c|}{TC} & 
\multicolumn{2}{|c|}{RS} & PP \\\cline{1-8}
$Re$ \quad & \quad 340 \quad & \quad 395 \quad & \quad 340 \quad & \quad 415 \quad & \quad 303 \quad & \quad 438 \quad & \quad 357 \\
$\lambda_\text{stream}$ \quad & \quad 110 \quad & \quad 110 \quad & \quad 145 \quad & \quad 95 \quad & \quad 71 \quad & \quad 106 \quad & \quad 103 \\
$\lambda_\text{span}$ \quad & \quad 83 \quad & \quad 52 \quad & \quad 70 \quad & \quad 35 \quad & \quad 24 \quad & \quad 36 \quad & \quad 45\\
$\lambda_z$ \quad & \quad 60 \quad & \quad 46 \quad & \quad 63 \quad & \quad 33 \quad & \quad 23 \quad & \quad 34 \quad & \quad 41\\
$\theta$ \quad & \quad $\:37^\circ$ \quad & \quad $\:25^\circ$ \quad & \quad $\:26^\circ$ \quad & \quad $\:20^\circ$ \quad & \quad $\:19^\circ$ \quad & \quad
$\:19^\circ$ \quad & \quad $\:24^\circ$\\
\cline{1-8}
\end{tabular}
\end{center}
\caption{Turbulent-laminar banded patterns in plane Couette (PC), 
Taylor-Couette (TC), rotor-stator (RS), and plane Poiseuille (PP) flow.
Parameters reported in \cite{Prigent_PhysD,Cros,Tsukahara}
are converted to a uniform Reynolds number based on the
average shear and half-gap, as described in the Appendix.
The two columns correspond to the values at the minimum 
and maximum Reynolds number reported.}
\label{tab:exp_obs}
\end{table}

Turbulent-laminar banded patterns have been observed in a number of shear
flows: plane Couette (PC) flow, Taylor-Couette (TC) flow, rotor-stator (RS)
flow (torsional Couette flow; the flow between differentially rotating disks)
and plane Poiseuille (PP) flow (channel flow).  Comparisons between these
flows are impeded by the fact that different conventions are used to
non-dimensionalise each of them.

In order to compare their observations in Taylor-Couette flow with those in
plane Couette flow, \cite*{Prigent_PhysD} generalise the Reynolds number used
in plane Couette flow $U=y/h$
%
%
by considering it as based on the shear and the half-gap:
\begin{equation}
Re^{PC} = \frac{(\text{Shear}^\text{PC})\;\text{(half-gap)}^2}{\nu} = 
\frac{\frac{U}{h} h^2}{\nu} = \frac{U h}{\nu}
\end{equation}
For flows whose shear is not constant, the average shear is used.
We also convert streamwise and spanwise wavelengths to total 
wavelength and angle of the pattern wavevector via
\begin{equation} 
\tan(\theta) = \frac{\lambda_\text{span}}{\lambda_\text{stream}} \qquad
\lambda_z = \lambda_\text{span} \cos(\theta)
\label{eq:convert}
\end{equation}
Table \ref{tab:exp_obs} presents the Reynolds numbers, wavelengths, and angles
for which turbulent-laminar patterns have been observed experimentally or
numerically. 
The subsections which follow explain how Table \ref{tab:exp_obs} was obtained
from the data in \cite{Prigent_PhysD,Cros,Tsukahara}.

\subsection{Taylor-Couette flow}

For Taylor-Couette flow between differentially rotating cylinders, 
the azimuthal and axial directions correspond to the streamwise and 
spanwise directions of plane Couette flow.
For cylinders of radius $r_i$ and $r_o$, rotating
at angular velocities $\omega_i$ and $\omega_o$ with $2h\equiv r_o-r_i$ and
$\eta\equiv r_i/r_o$, the shear averaged over the gap is
\begin{equation}
\langle \text{Shear}^\text{TC}\rangle = 
\frac{r_i\omega_i - \eta r_o\omega_o}{(1+\eta)h}
\end{equation}
leading to the Reynolds number:
\begin{equation}
Re^\text{TC} \equiv \frac{r_i \omega_i - \eta r_o \omega_o}{(1+\eta)h} \;
\frac{h^2}{\nu} \approx \frac{Re_i-Re_o}{4\nu}
\end{equation}
where the last approximate equality corresponds to exact counter-rotation
($\omega_o=-\omega_i$) and the narrow gap limit ($\eta \rightarrow 1$), and
$R_i$, $R_o$ are the conventionally defined inner and outer Reynolds numbers,
\eg\ $Re_i\equiv 2h r_i\omega_i/\nu$.  
The wavelengths and Reynolds numbers observed in Taylor-Couette and plane
Couette flow are compared in Figure 5 of \cite{Prigent_PhysD}.

\subsection{Torsional Couette flow}

The laminar profile for torsional Couette flow between a rotating and a
stationary disk (rotor-stator flow) is
\begin{equation}
\bu=\be_\theta \;\frac{\omega \,r \,z}{h}
\end{equation}
and the Reynolds number based on axial shear and half-gap is
\begin{equation}
Re^\text{RS}=\frac{\omega r}{h} \;\frac{h^2}{4\nu} = \frac{\omega r h}{4\nu}
\label{eq:Re_RS}\end{equation}
For $m$ spirals, the azimuthal wavelength in units of the half-gap is
\begin{equation}
\lambda^\text{RS}_\text{stream} = \frac{2\pi r}{m h/2}=\frac{4\pi r}{m h}
\label{eq:lambda_RS}\end{equation}
Turbulent spiral patterns which are rather regular occur for a range of
angular velocities and radii.  In their figures 12, 16 and 18, \cite{Cros}
focus particularly on the radius and gap:
\begin{equation}
r = 0.8\times 140\,\text{mm}=11.2\,\text{cm} \qquad
h=0.22\,\text{cm}
\label{eq:rh_RS}\end{equation}
The highest and lowest rotation rates for which turbulent spirals are seen are
\begin{subeqnarray}
\omega = 68\,\text{rev/min} = 
7.12 \,\text{rad/sec} &&\qquad\text{with } m=6 \\
\label{eq:omegahi_RS}
%
%
%
%
%
\omega = 47\,\text{rev/min} = 
4.92 \,\text{rad/sec} &&\qquad\text{with } m=9
\label{eq:omegalo_RS}
\end{subeqnarray}
%
%
%
Substituting \eqref{eq:rh_RS}-\eqref{eq:omegalo_RS} and the viscosity
$\nu=10^{-2}\,$cm$^2$/sec of water into \eqref{eq:Re_RS}-\eqref{eq:lambda_RS}
leads to the values shown in Table \ref{tab:exp_obs}.  The pitch angle of the
spirals remains approximately constant at $19^\circ$.  We use
\eqref{eq:convert} to calculate $\lambda_\text{span}$ and $\lambda_z$,
neglecting the variation in radius.

\subsection{Plane Poiseuille flow}

Figure 14 of \cite*{Tsukahara} shows a visualisation from a direct
numerical simulation of plane Poiseuille (PP) flow in a channel
with domain and Reynolds number
\begin{equation}
L_\text{stream}\times L_y\times L_\text{span} = 
51.2\;\delta \times 2\;\delta \times 22.5\;\delta
\qquad
Re_c \equiv \frac{u_c\delta}{\nu} = 1430
\end{equation}
where $u_c$ is the centerline velocity.
The domain contains a single wavelength of an oblique turbulent-laminar banded
pattern oriented at $\theta=24^\circ$ to the streamwise direction.
(Both the wavelength and the angle are dictated by the computational domain.)
Following \cite{Waleffe_03}, we view the Poiseuille profile in the
half-channel $[-\delta,0]$, over which the shear has one sign, as comparable
to the Couette profile in the channel $[-h,h]$, and thus take $\delta/2$ as
the unit of length, rather than $\delta$.  The shear is obtained by
averaging over $[-\delta,0]$:
\begin{equation}
\langle\text{Shear}^\text{PP}\rangle 
= \left\langle \frac{du}{dy} \right\rangle 
= \frac{u_c}{\delta}
\end{equation}
For the Reynolds number based on the average shear and half-gap, we obtain
\begin{equation}
Re^\text{PP}=\frac{\langle\text{Shear}^\text{PP}\rangle 
\;(\text{half-gap})^2}{\nu} = 
\frac{\frac{u_c}{\delta}\frac{\delta^2}{4}}{\nu} = 
\frac{u_c\delta}{4\nu} = \frac{Re_c}{4\nu} = \frac{1430}{4}=357.5
\end{equation}

\bibliographystyle{jfm}
\bibliography{bands}

\end{document}